\documentclass[preprint,aps ,nofootinbib]{revtex4}
\pdfoutput=1
\usepackage{graphicx}%Include figure files
\usepackage{epsfig}
\usepackage{amsmath}
\usepackage{amsfonts}
\usepackage{amssymb}
\usepackage{color}%
\usepackage{dcolumn}% Align table columns on decimal point
\usepackage{slashed}
\usepackage{hyperref}
\usepackage{enumerate}

\def\arXiv#1{\href{http://arxiv.org/abs/#1}{arXiv:#1}}
\def\arXiv#1#2{\href{http://arxiv.org/abs/#1}{arXiv:#1}}
\def\arXivid#1#2{\href{http://arxiv.org/abs/#1/#2}{#1/#2}}
\def\doi#1#2{\href{http://doi.org/#1}{#2}}

\providecommand{\U}[1]{\protect\rule{.1in}{.1in}}

\newcommand{\f}{\begin{equation}}
\newcommand{\ff}{\end{equation}}
\newcommand{\fa}{\begin{eqnarray}}
\newcommand{\ffa}{\end{eqnarray}}

\begin{document}
\title{Holographic Fermions in Striped Phases}
\author{Sera Cremonini$^a$}
\email{cremonini@lehigh.edu}
\author{Li Li$^{a}$}
\email{lil416@lehigh.edu}
\author{Jie Ren$^{b}$}
\email{jie.ren@mail.huji.ac.il}
\affiliation{$^a$ Department of Physics, Lehigh University, Bethlehem, PA, 18018, USA.\\
$^b$Racah Institute of Physics, The Hebrew University of Jerusalem, 91904, Israel.
 }

\begin{abstract}

We examine the fermionic response in a holographic model of a low temperature striped phase,
working for concreteness with the setup we studied in~\cite{Cremonini:2016rbd,Cremonini:2017usb}, in which 
a U(1) symmetry and translational invariance are broken spontaneously at the same time.
We include an ionic lattice that breaks translational symmetry explicitly in the UV of the theory. 
Thus, this construction realizes spontaneous crystallization on top of a background lattice.
We solve the Dirac equation for a probe fermion in the associated background geometry
using numerical techniques, and explore the interplay between spontaneous and explicit breaking of translations.
We note that in our model the breaking of the 
U(1) symmetry doesn't play a role in the analysis of the fermionic spectral function.
We investigate under which conditions a Fermi surface can form 
and focus in particular on how the ionic lattice affects its structure. 
When the ionic lattice becomes sufficiently strong the spectral weight peaks broaden, 
denoting a gradual disappearance of the Fermi surface along the symmetry breaking direction. 
This phenomenon occurs even in the absence of spontaneously generated stripes.
The resulting Fermi surface appears to consist of detached segments reminiscent of Fermi arcs.
%%%%%

\end{abstract}
\maketitle

\tableofcontents

\newpage
\section{Introduction}

Holographic methods provide a theoretical laboratory for probing exotic phases of matter that lie outside the paradigm of Laundau's Fermi liquid theory. Within this program, in recent years we have seen many studies of fermionic response in strongly interacting systems, using the techniques of holography (see~\cite{Henningson:1998cd, Mueck:1998iz, Lee:2008xf, Liu:2009dm, Cubrovic:2009ye, Faulkner:2009wj} for pioneering works and \emph{e.g.}~\cite{Iqbal:2011ae} for a review). In particular, holographic spectral functions can be compared with measurements of Angle-Resolved Photoemission Spectroscopy (ARPES) or Scanning Tunneling Microscopy (STM) experiments,
thus potentially providing a crucial test for the applications of holography to real materials. 
The majority of these studies thus far has focused on homogeneous cases which respect translational invariance -- they involve gravitational constructions whose bulk metrics
depend only on the holographic radial direction.  However, real materials such as the copper oxides exhibiting high temperature superconductivity~\cite{Jan:2015} 
are characterized by very strong lattice potentials which break translational symmetry.

The scarcity of holographic analyses of fermionic spectral functions in the presence of lattices can be traced to the difficulties 
that arise when examining the associated systems of non-linear partial differential equations. 
Studies of homogeneous holographic lattices, which simulate the effects of translational symmetry breaking while retaining the homogeneity of the spacetime geometry~\cite{Donos:2012js,Donos:2013eha,Andrade:2013gsa}, have proven successful at obtaining finite conductivities in holographic models of metallic phases, by reproducing the expected Drude physics. 
Nevertheless, as demonstrated by the authors of~\cite{Bagrov:2016cnr}, to describe holographically lattices relevant for condensed matter applications, one needs to consider periodic lattices rather than homogeneous ones.
The first such study appeared in~\cite{Liu:2012tr}, where a lattice structure was encoded through a perturbatively small periodic modulation of the chemical potential, neglecting the backreaction on the metric. 
This analysis, performed in a weak potential limit, was then generalized to the fully backreacted case in~\cite{Ling:2013aya}, where the periodic gravitational backgrounds were constructed numerically. 
Certain interesting characteristics were identified, such as an anisotropic Fermi surface and the appearance of a band gap at the Brillouin zone boundary. However, these analyses only focus on cases in which the lattice periodicity is put in by hand and is irrelevant in the infrared.

The purpose of this paper is to investigate the fermionic response in a quantum phase of matter 
in which a U(1) symmetry and translational invariance are broken spontaneously and at the same time, resulting in a concrete realization of interwined orders. 
In particular, we work within a holographic bottom-up model \cite{Cremonini:2016rbd,Cremonini:2017usb} of a striped superconductor
which realizes certain key features of pair density wave (PDW) order, and more generically of phases in which 
charge density wave (CDW) and superconducting (SC) orders co-exist. 
Compelling experimental evidence of a PDW has accumulated in cuprate superconductors~\cite{CMPDW1,CMPDW2,CMPDW3}, and there is
also computational evidence suggesting that it might be a robust feature of strongly correlated electron systems \cite{FKT}. For condensed matter models that have been proposed to describe the properties of such novel strongly coupled phases we refer the reader to \emph{e.g.}~\cite{Baruch:2008,Berg:2009,Lee:2014,Garrido:2015}.
With an available holographic model for PDWs, it is valuable to investigate the structure of the associated fermionic spectral function, and in particular examine under what conditions a Fermi surface will form, whether it will exhibit a gap and what controls its properties. 
However, while we are building on our previous work on PDW order, we stress that our focus in this paper is on the effects of broken translational symmetry on the Fermi surface properties, and \emph{not} on the interaction between the fermion and the superconducting condensate. 
We comment on ways to examine such interactions in the Conclusions, and leave a detailed analysis to future work.

In particular, we will explore the fermionic spectral function by solving numerically the Dirac equation~\eqref{diraceom} in a low temperature PDW phase, in two different cases. 
We will work first
in a pure PDW phase in which a U(1) symmetry and translational invariance are both broken spontaneously by the same underlying mechanism. 
We will then add an ionic lattice which breaks translations explicitly in the UV of the theory, 
so that the final construction %%%%
realizes spontaneous crystallization in the presence of a background lattice.
As we will see, the formation of the Fermi surface will require a sufficiently large value of the fermionic charge, as already known from the literature.
A more interesting feature involves the shape and structure of the Fermi surface (including the presence and size of a gap\,\footnote{We stress that this is not a superconducting gap, but due to the broken translational invariance and the periodic modulation of the background.}),  which will be highly sensitive to the strength of broken translational 
invariance. Intriguingly, we will find a gradual disappearance of the Fermi surface as the strength of the lattice becomes too large, as discussed in detail in Section \ref{Numerics}.
We expect the Fermi surface features we identify to be widely applicable to striped superconducting phases as well as other spatially modulated or striped phases, and not just to the physics of a PDW. We will present further evidence by considering a model with only an ionic lattice.

The structure of the paper is the following. 
The holographic model we will work with is introduced in Section \ref{SetupSection}, and its gravitational solutions are included in Section \ref{backreaction}.
The Dirac equation and spectral function are discussed in Section \ref{Greenfunction}, and the numerical analysis is presented in Section \ref{Numerics}.
We conclude in Section \ref{Conclusions} with a summary of results and future directions.
The analysis of the energy distribution of the spectral function is relegated to Appendix~\ref{appA}
 and further details of the numerical analysis to Appendix~\ref{appB}.

%%%%%%%%%%%%%%%%%
%%%%%%%%%%%%%%%%%
\section{Gravity Setup}
\label{SetupSection}
%%%%%%%%%%%%%%%%%
%%%%%%%%%%%%%%%%%

The holographic model we work with involves two real scalar fields $\chi$ and $\theta$ coupled to two abelian vector fields $A_\mu$ and $B_\mu$,
\begin{eqnarray}
S&=&\frac{1}{2\kappa_N^2}\int d^{4}x \sqrt{-g} \left[\mathcal{R}+\frac{6}{L^2}+\mathcal{L}_{m}\right], \nonumber \\
\mathcal{L}_{m} &=& -\frac{1}{2}\partial_{\mu}\chi \partial^{\mu}\chi-\frac{Z_A(\chi)}{4}F_{\mu\nu}F^{\mu\nu}-\frac{Z_B(\chi)}{4}\tilde{F}_{\mu\nu}\tilde{F}^{\mu\nu}-\frac{Z_{AB}(\chi)}{2}F_{\mu\nu}\tilde{F}^{\mu\nu}\nonumber\\
&&-\mathcal{K}(\chi)(\partial_\mu\theta-q_A A_\mu-q_B B_\mu)^2
%-\frac{m_v^2}{2}B_\mu B^\mu
-V(\chi)\,,
\label{actions}
\end{eqnarray}
with $F_{\mu\nu}=\partial_\mu A_\nu-\partial_\nu A_\mu$ and $\tilde{F}_{\mu\nu}=\partial_\mu B_\nu-\partial_\nu B_\mu$ their respective field strengths.
This model was studied first in~\cite{Cremonini:2016rbd,Cremonini:2017usb} to realize the idea of intertwined orders in holography, through the spontaneous breaking of both translational invariance and a U(1)  symmetry at the same time.  
The scalar $\chi$ generically couples to both vectors\,\footnote{This model falls within the generalized class of holographic superconductors via St\"ukelberg term (known as ``Josephson action" in the condensed matter literature), see~\emph{e.g.}~\cite{Aprile:2009ai,Aprile:2010yb,Cai:2012es,Kiritsis:2015hoa}.}. Depending on the choice of parameters $q_A$ and $q_B$, the bulk theory can describe different striped quantum phases. 
As shown in detail in~\cite{Cremonini:2016rbd,Cremonini:2017usb}, the case with $q_B=0$ and $q_A\neq 0$ enables us to mimic certain features of PDW order, while the case with $q_B\neq0$ and $q_A\neq 0$ corresponds to a state with coexisting superconducting and CDW orders, in which the scalar condensate has a uniform component. A pure CDW state without U(1) symmetry breaking can be obtained by setting $q_A=q_B=0$ and consistently truncating $\theta$~\cite{Donos:2013gda,Ling:2014saa}.

In the present paper we would like to investigate the fermionic response associated with these spatially modulated phases, and possibly identify any generic signature they may possess. 
To this end we consider the bulk action for a probe Dirac fermion  $\zeta$ with charge $q$ and mass $m$,
\begin{eqnarray}\label{actionfermion}
S_{D}=i\int d^{4}x \sqrt{-g}\,\overline{\zeta}\left(\slashed{D}-m\right)\zeta\,,
\end{eqnarray}
with $\slashed{D}=\Gamma^{\underline{a}}e_{\underline{a}}^\mu\,(\partial_{\mu}+\frac{1}{4}(\omega_{\underline{ab}})_{\mu}\Gamma^{\underline{ab}}-iqA_{\mu})$ and $\bar{\zeta}=\zeta^{\dag} \Gamma^{\underline{t}}$.
Here $(\underline{a},\underline{b})$ denote the tangent indices, $\Gamma^{\underline{a}}$ are gamma matrices with $\Gamma^{\underline{bc}}=\frac{1}{2}[\Gamma^{\underline{b}},\Gamma^{\underline{c}}]$, and $e_{\underline{a}}^\mu$ is the vielbein with $(\omega_{\underline{ab}})_{\mu}=e_{\underline{a}\nu}\nabla_\mu e_{\underline{b}}^\nu$ the associated spin connection. The correlation function for the fermionic operator of the strongly coupled dual field theory is then obtained by solving the bulk Dirac equation.

%%%%%%%%%%%%%%%%%%%%%%%%%%%%
%%%%%%%%%%%%%%%%%%%%%%%%%%%%
%%%%%%%%%%%%%%%%%%%%%%%%%%%%
\section{The Striped Solutions}
\label{backreaction}
%%%%%%%%%%%%%%%%%%%%%%%%%%%%
%%%%%%%%%%%%%%%%%%%%%%%%%%%%
%%%%%%%%%%%%%%%%%%%%%%%%%%%%

In this section we construct the non-linear solutions corresponding to the spatially modulated black branes. We take the couplings in~\eqref{actions} to be given by
\begin{equation}\label{couplingnumerics}
\begin{split}
& Z_A(\chi)=1+2\chi^2,\quad Z_B(\chi)=1,\quad Z_{AB}(\chi)=-2.34\, \chi\,, \\
& V( \chi)=-\frac{1}{L^2}\chi^2,\quad \mathcal{K}(\chi)=\frac{1}{2}\chi^2\,,
\end{split}
\end{equation}
so that the scalar operator dual to $\chi$ has dimension $\Delta=2$. As shown in our previous work~\cite{Cremonini:2016rbd,Cremonini:2017usb}, striped order will develop spontaneously below a 
certain critical temperature, with an intrinsic wavelength $k$ which depends on the details of the theory. As a typical example, throughout we are going to focus on the $k/\mu$ = 1 branch, with the corresponding critical temperature being $T_c/\mu = 0.016$.\,\footnote{We point out that this branch may not be thermodynamically preferred, but this does not affect our discussion on the fermionic spectral function.} We will work in the grand canonical ensemble by fixing the chemical potential and typically setting it to $\mu=1$.

We focus on the uni-directional striped solutions and employ the DeTurck trick~\cite{Headrick:2009pv} to solve the resulting system. To implement the DeTurck method, one needs to choose a reference metric, for which we use the AdS Reissner-Nordstr\"{o}m (AdS-RN) black brane
\begin{equation}\label{RNadsz}
\begin{split}
&ds^2=\frac{r_h^2}{L^2 (1-z^2)^2}\left[-F(z)dt^2+\frac{4 z^2 L^4}{r_h^2 F(z)}dz^2+(dx^2+dy^2)\right]\,,\\
&F(z)=z^2\left[2-z^2+(1-z^2)^2-\frac{L^2 \mu^2}{4 r_h^2}(1-z^2)^3\right],\quad A_t=\mu\, z^2\,.
\end{split}
\end{equation}
Note that we are working in the coordinate system\footnote{One can switch to the standard holographic coordinate used in the literature via $z^2=1-r_h/r$ (see, for example, the AdS-RN metric in Eq. (2.9) of our previous paper~\cite{Cremonini:2017usb}). In the more standard coordinate $r$, the IR regularity condition requires all functions to have an analytic expansion in powers of $(r-r_h)$, which corresponds to the $z^2$ expansion in our present coordinate. As we will see, the $z$ coordinate is convenient for solving the striped geometry numerically.}  in which the horizon is located at $z=0$ and the AdS boundary at $z=1$. Here $r_h$ is a free parameter that determines the black hole temperature
\begin{equation}
\label{temp}
T=\frac{r_h}{4\pi}\left[\frac{3}{L^2}-\frac{\mu^2}{4r_h^2}\right]\,.
\end{equation}
We adopt the following ansatz for the striped  black brane,
\begin{eqnarray}\label{ansatzbh}
&&ds^2=\frac{r_h^2}{L^2 (1-z^2)^2}\left[-F(z)Q_{tt} \, dt^2+\frac{4 z^2 L^4 Q_{zz}}{r_h^2 F(z)}\, dz^2+Q_{xx}(dx-2 z(1-z^2)^2Q_{xz}dz)^2+Q_{yy}\, dy^2\right]\,,\nonumber \\
&&\chi=(1-z^2)\phi \, , \qquad A_t=\mu\, z^2 \alpha, \qquad B_t=z^2 \beta\,,
\end{eqnarray}
where the eight functions $(\phi,\alpha,\beta,Q_{tt},Q_{zz},Q_{xx},Q_{yy}, Q_{xz})$ depend on  $z$ and the spatial coordinate $x$ along which translational symmetry will be broken.
One recovers the AdS-RN solution~\eqref{RNadsz} by choosing their background values to be $\phi=\beta=Q_{xz}=0$, $\alpha=Q_{tt}=Q_{zz}=Q_{xx}=Q_{yy}=1$.

This ansatz results in a system of equations of motion involving eight PDEs in the variables $z$ and $x$. Here we discuss only briefly the numerical procedure we used, but we refer the reader to~\cite{Cremonini:2017usb} for further details. We adopt the pseudo-spectral collocation approximation to convert the PDEs into non-linear algebraic
equations, by adopting Fourier discretization in the $x$ direction and Chebyshev polynomials in the $z$ direction. The resulting system is then solved using a Newton-Raphson method with appropriate boundary conditions. 

Since we seek solutions with a regular horizon at $z=0$, we require all functions to depend on $z^2$ smoothly. Therefore, one can impose Neumann boundary conditions of the type $\partial_z \phi(0,x)=0$, and similarly for the remaining components at the horizon. There is an additional Dirichlet boundary condition $Q_{tt}(0,x)=Q_{zz}(0,x)$, which ensures that the temperature of the black brane~\eqref{ansatzbh} is the same as~\eqref{temp}.

On the other hand, the UV boundary conditions are slightly more involved. For the pure PDW phase in the absence of a background lattice, both the U(1) symmetry and the spatially translational invariance are broken spontaneously.
To ensure spontaneous symmetry breaking we take all sources to vanish, and in addition fix the metric to be asymptotically AdS at the UV boundary $z=1$,
\begin{equation}
\begin{split}
&\phi(1,x)=\beta(1,x)=Q_{xz}(1,x)=0\,,\\
&Q_{tt}(1,x)=Q_{zz}(1,x)=Q_{xx}(1,x)=Q_{yy}(1,x)=\alpha(1,x)=1\,.
\end{split}
\end{equation}

Clearly, it is also of interest to study spontaneous holographic crystallization in the presence of a background lattice. A simple way to do so is to introduce an ionic lattice which breaks the translational symmetry explicitly. In this case we expect the Goldstone mode due to the spontaneously broken translational invariance to acquire a mass and become pinned. 
The ionic potential can be introduced by imposing a spatially varying boundary condition for the chemical potential, 
\begin{equation}\label{lattice}
\mu(x)=A_t(1,x)=\mu[1+a_0 \cos(p\, x)]\,,
\end{equation}
i.e. a uni-directional single harmonic potential with wavevector $p$ and relative amplitude $a_0$. 
We emphasize that we are working with a system that has two wavevectors -- the intrinsic scale $k$ (associated with the spontaneous breaking of translations) and the lattice scale $p$ put in by hand
(associated with explicit symmetry breaking). 
One anticipates that when these two scales are sufficiently close together there will be a ``commensurate lock-in" of the spontaneous crystal, resulting in additional stability~\cite{Andrade:2017leb,Andrade:2017ghg}.  In the present study we focus on the case with $p=2k$. Here $k$ is the intrinsic wavelength associated with the spontaneous modulations in the absence of a lattice, and the factor of 2 is introduced to match the period of the charge density oscillations\,\footnote{Note that, as shown in our previous work~\cite{Cremonini:2016rbd,Cremonini:2017usb}, the condensate of the scalar in the PDW phase induces a sub-leading modulation of the charge density with a frequency $2k$. }. 
The fact that the two periods coincide means that the ionic lattice is commensurate with the charge density wave in the PDW phase.

\begin{figure}[ht!]
\begin{center}
\includegraphics[width=.46\textwidth]{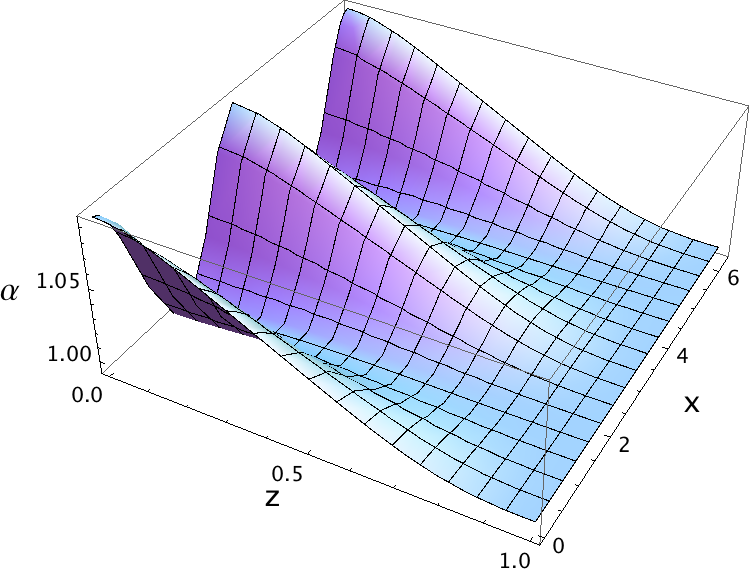}\quad
\includegraphics[width=.46\textwidth]{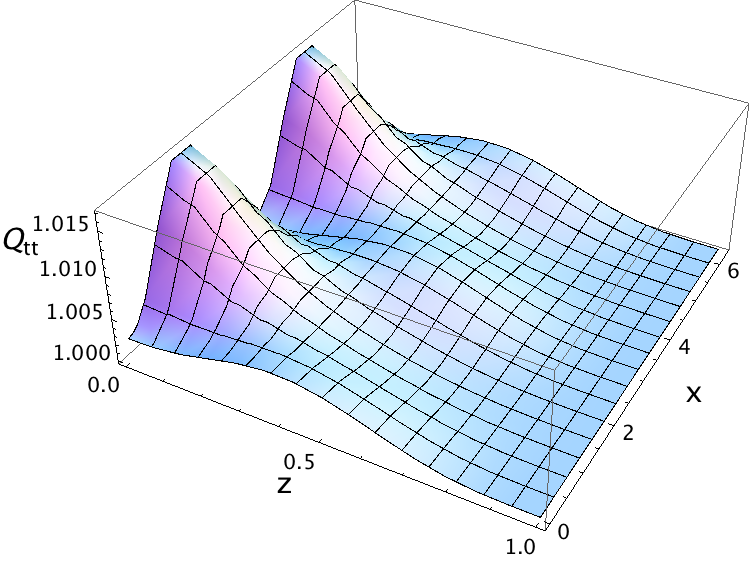}
\includegraphics[width=.46\textwidth]{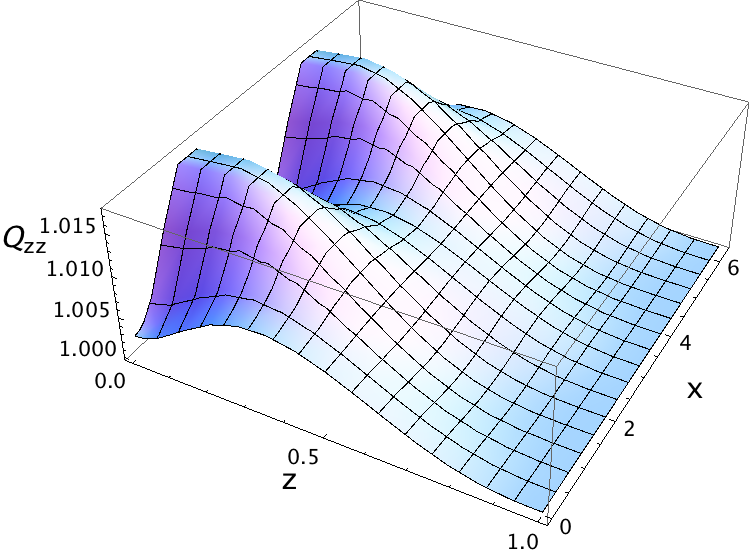}\quad
\includegraphics[width=.46\textwidth]{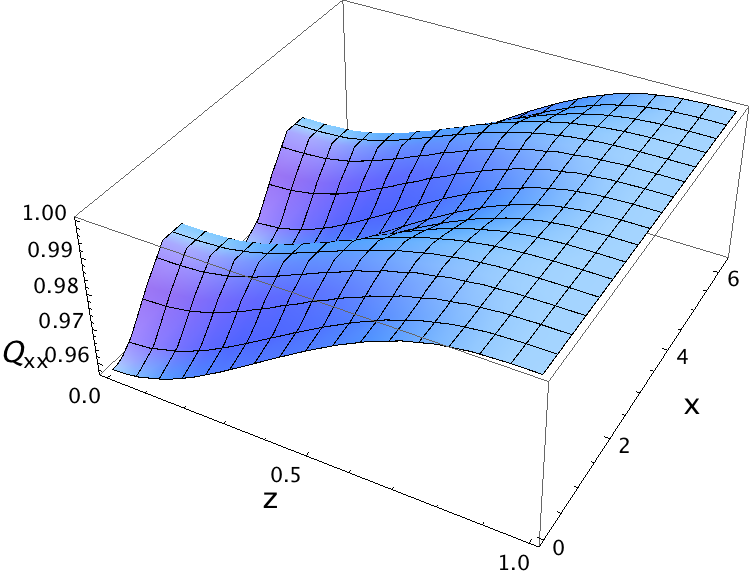}
\caption{Background configurations for the gauge field $\alpha$ and metric fields $Q_{tt}, Q_{zz}, Q_{xx}$ at $T=0.01426$ and $k=1$, for the PDW phase without ionic lattice. The horizon is located at $z=0$ and the AdS boundary at $z=1$. We have set $L=1/2$.}
\label{fig:geometry}
\end{center}
\end{figure}

Profiles for the bulk fields corresponding to the pure PDW phase are shown in figure~\ref{fig:geometry}. 
Notice that since we have not turned on any source, all bulk configurations are homogeneous at the UV boundary $z=1$. It is clear that the spatial modulations are imprinted near the horizon at $z=0$, and decrease in overall magnitude as the AdS boundary is approached. This is due to the fact that in our theory the striped feature is a relevant deformation of the UV field theory and is strongest in the IR.

\begin{figure}[ht!]
\begin{center}
\includegraphics[width=.44\textwidth]{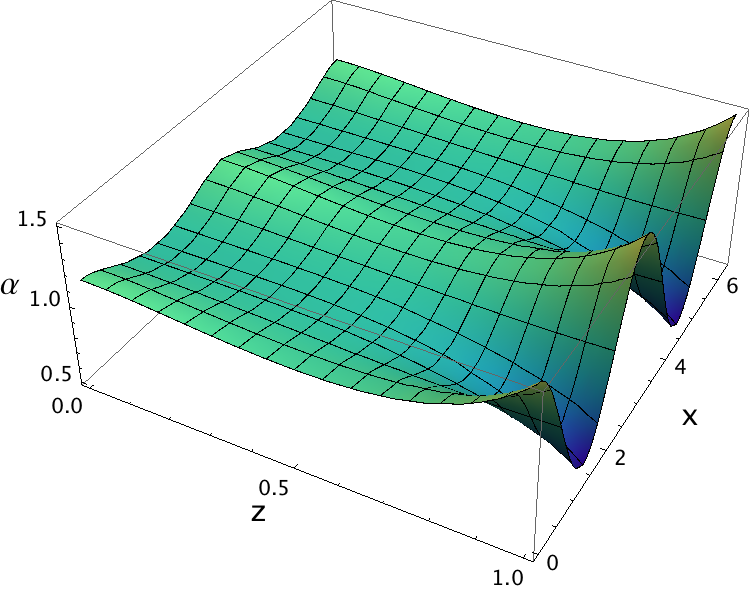}\quad
\includegraphics[width=.45\textwidth]{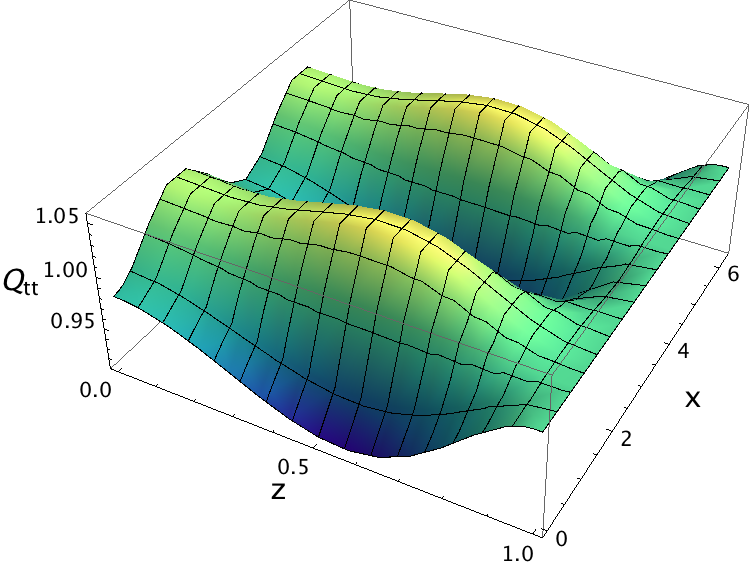}
\includegraphics[width=.45\textwidth]{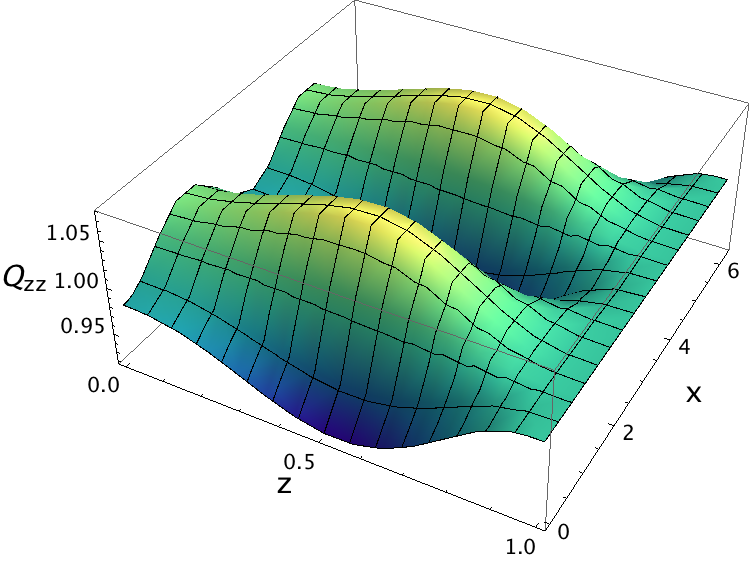}\quad
\includegraphics[width=.45\textwidth]{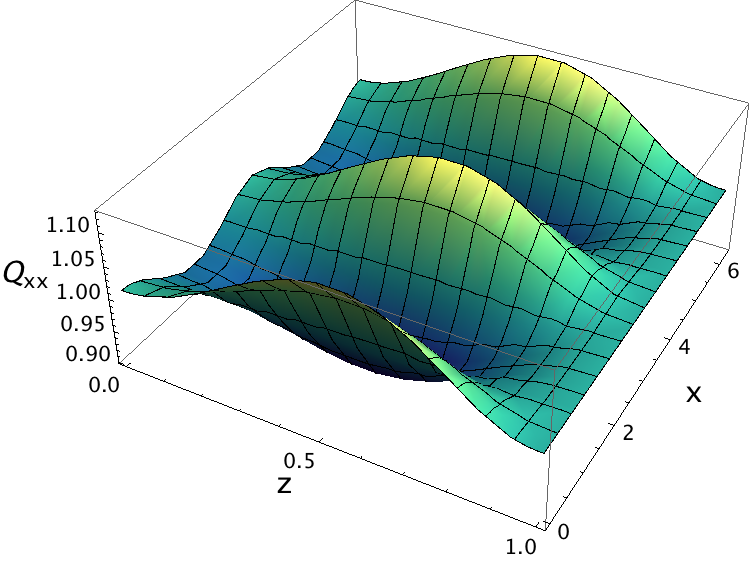}
\caption{Background configurations for the gauge field $\alpha$ and metric fields $Q_{tt}, Q_{zz}, Q_{xx}$ at $T=0.01426$ and $k=1$, for the PDW phase with the ionic lattice $\mu(x)=1+0.5\cos(2\,x)$. The horizon is located at $z=0$ while the AdS boundary at $z=1$. We have set $L=1/2$.}
\label{fig:geometryIonic}
\end{center}
\end{figure}

Representative profiles for the bulk fields in the presence of the ionic lattice are shown in figure~\ref{fig:geometryIonic}. This case corresponds to the holographic crystallization in the presence of an external periodic potential, as clearly visible from the profile of the gauge field $\alpha$ at the UV boundary $z=1$.  The bulk modulations are due to the following two mechanisms. 
One is the spontaneous translational symmetry breaking which is a relevant deformation of the UV field theory, and the other is the explicit UV lattice which is instead an irrelevant deformation. 
As a consequence, the strength of the striped oscillations, clearly visible from the metric fields of figure~\ref{fig:geometryIonic}, shows a non-monotonic behavior along the radial direction, encoding a more complicated RG flow from the UV to the IR.
%

%%%%%
%%%%%
\section {The Dirac Equation and Spectral Function}
\label{Greenfunction}
%%%%%
%%%%%
%%%%%

We are now ready to discuss the Dirac equation that is used to compute the retarded Green's function for the fermionic operator of the strongly coupled field theory. 
The bulk Dirac equation obtained from~\eqref{actionfermion} reads
\begin{eqnarray}\label{diraceom}
\left[\Gamma^{\underline{a}} \, e_{\underline{a}}^\mu\,(\partial_{\mu}+\frac{1}{4}(\omega_{\underline{ab}})_{\mu}\Gamma^{\underline{ab}}-iqA_{\mu})-m\right]\zeta=0\,.
\end{eqnarray}
The vielbein and gamma matrices can be chosen in many different ways. Note that for the background geometry~\eqref{ansatzbh} the horizon is located at $z=0$, while the AdS boundary at $z=1$. We use the following vielbein,
\begin{equation}
\begin{split}
&
e_{\underline{t}}=\frac{L}{r_h}\frac{(1-z^2)}{\sqrt{F(z)Q_{tt}(z,x)}}\frac{\partial}{\partial t}\,,~~\\
&
e_{\underline{x}}=\frac{L}{r_h}\frac{(1-z^2)}{\sqrt{Q_{xx}(z,x)}}\frac{\partial}{\partial x},\quad\quad\quad\quad\quad\quad\quad
e_{\underline{y}}=\frac{L}{r_h}\frac{(1-z^2)}{\sqrt{Q_{yy}(z,x)}}\frac{\partial}{\partial y}\,,~~\\
&
e_{\underline{z}}=\frac{(1-z^2)^3 Q_{xz}(x,z)}{L}\sqrt{\frac{F(z)}{Qzz(z,x)}}\frac{\partial}{\partial x}+\frac{(1-z^2)}{2Lz}\sqrt{\frac{F(z)}{Qzz(z,x)}}\frac{\partial}{\partial z}\,,
\end{split}
\end{equation}
and consider a basis of gamma matrixes given by
\begin{equation}\label{Gamma}
\begin{split}
& \Gamma^{\underline{t}} = \left[ \begin{array}{cc}
 i \sigma^1  & 0  \\
0 & i \sigma^1
\end{array} \right], \quad
\Gamma^{\underline{x}} = \left[ \begin{array}{cc}
-\sigma^2  & 0  \\
0 & \sigma^2
\end{array} \right], \\
& \Gamma^{\underline{y}} = \left[ \begin{array}{cc}
 0  & \sigma^2  \\
\sigma^2 & 0
\end{array} \right],\quad\;\;
\Gamma^{\underline{z}} = \left[ \begin{array}{cc}
-\sigma^3  & 0  \\
0 & -\sigma^3
\end{array} \right]\,,
\end{split}
\end{equation}
where $(\sigma^1, \sigma^2, \sigma^3)$ are Pauli matrixes.
We proceed by a redefinition of $\zeta$ via
\begin{equation}
\zeta=\left(\frac{L(1-z^2)}{r_h}\right)^{3/2}(FQ_{tt}Q_{xx}Q_{yy})^{-1/4}\left[\begin{matrix} \Psi_{1} \cr  \Psi_{2} \end{matrix}\right]\,,
\end{equation}
with each $\Psi_\alpha$, $\alpha=1,2$ a two-component spinor. In our striped phase~\eqref{ansatzbh} all background configurations are spatially modulated in the $x$ direction, 
with the periodicity fixed by the Umklapp wavevector $K$. In contrast to the homogeneous case, the different momentum modes are no longer independent, and the Green's function will therefore have non-zero value for operators with momenta that differ by a lattice vector. According to the Bloch theorem, we then adopt the following expansion~\cite{Liu:2012tr},
\begin{equation}\label{bloch}
\Psi_{\alpha}=\int \frac{d\omega d k_x d k_y}{2\pi}\sum_{n=0,\pm1,\pm2,\cdots} \mathcal{F}_\alpha^{(n)}(z,\omega,k_x, k_y)e^{-i\omega t+i(k_x+n K)x+i k_y y}\,.
\end{equation}
Here $k_x\in[-\frac{K}{2},\frac{K}{2}]$ and $n$ characterizes the momentum level or Brillouin zone.
The Bloch expansion~\eqref{bloch} can also be written as
\begin{equation}\label{blochx}
\Psi_{\alpha}=\int \frac{d\omega d k_x d k_y}{2\pi}\mathcal{F}_\alpha(z,x,\omega, k_y)e^{-i\omega t+ik_xx+i k_y y}\,,
\end{equation}
where
\begin{equation}
\mathcal{F}_\alpha(z,x,\omega, k_y)=\sum_{n=0,\pm1,\pm2,\cdots} \mathcal{F}_\alpha^{(n)}(z,\omega,k_x, k_y)e^{i n K x}\,,
\end{equation}
is a periodic function of $x$ with periodicity $2\pi/K$, i.e., $\mathcal{F}_\alpha(z,x,\omega, k_y)=\mathcal{F}_\alpha(z,x+\frac{2\pi}{K},\omega, k_y)$.

For further convenience, we decompose $\mathcal{F}_\alpha$ in the following way\,\footnote{For the sake of brevity, we will drop the parameters $(\omega, k_x, k_y)$ in most expressions from now on.}
\begin{eqnarray}
\mathcal{F}_{\alpha}(z,x) = \left[ \begin{matrix} \mathcal{A}_{\alpha}(z,x) \cr  \mathcal{B}_{\alpha}(z,x) \end{matrix}\right],\quad \alpha=1, 2\,.
\end{eqnarray}
The Dirac equation~\eqref{diraceom} can then be expressed as
\begin{eqnarray} \label{Dirac1}
\left(\partial_{z}+2z(1-z^2)^2Q_{xz}\partial_x+\Pi_{1}\pm\frac{2 mL z}{1-z^2}\sqrt{\frac{Q_{zz}}{F}}\right)
\left[ \begin{matrix} \mathcal{A}_{1} \cr  \mathcal{B}_{1} \end{matrix}\right]
&\mp& \Pi_{2}\left[ \begin{matrix} \mathcal{B}_{1} \cr  \mathcal{A}_{1} \end{matrix}\right]\\\nonumber
- i\frac{2L^2z}{r_h}\sqrt{\frac{Q_{zz}}{F Q_{xx}}}(\partial_x+\Pi_3)\left[ \begin{matrix} \mathcal{B}_{1} \cr  \mathcal{A}_{1} \end{matrix}\right]
&-&\frac{2L^2z}{r_h}\sqrt{\frac{Q_{zz}}{F Q_{yy}}}k_y \left[ \begin{matrix} \mathcal{B}_{2} \cr  \mathcal{A}_{2} \end{matrix}\right]
=0,
\end{eqnarray}
\begin{eqnarray} \label{Dirac2}
\left(\partial_{z}+2z(1-z^2)^2Q_{xz}\partial_x+\Pi_{1}\pm\frac{2 mL z}{1-z^2}\sqrt{\frac{Q_{zz}}{F}}\right)
\left[ \begin{matrix} \mathcal{A}_{2} \cr  \mathcal{B}_{2} \end{matrix}\right]
&\mp&\Pi_{2}\left[ \begin{matrix} \mathcal{B}_{2} \cr  \mathcal{A}_{2} \end{matrix}\right]\\\nonumber
+ i\frac{2L^2z}{r_h}\sqrt{\frac{Q_{zz}}{F Q_{xx}}}(\partial_x+\Pi_3)\left[ \begin{matrix} \mathcal{B}_{2} \cr  \mathcal{A}_{2} \end{matrix}\right]
&-&\frac{2L^2z}{r_h}\sqrt{\frac{Q_{zz}}{F Q_{yy}}}k_y \left[ \begin{matrix} \mathcal{B}_{1} \cr  \mathcal{A}_{1} \end{matrix}\right]
=0,
\end{eqnarray}
where
\begin{equation}
\begin{split}
\Pi_1&=z(1-z^2)^2[2 i \,k_x\, Q_{xz}(z,x)+\partial_x Q_{xz}(z,x)]\,,\\
\Pi_2&=\frac{2L^2z}{r_h F(z)}\sqrt{\frac{Q_{zz}(z,x)}{Q_{tt}(z,x)}}[\omega+q\mu z^2\,\alpha(z,x)]\,,\\
\Pi_3&=i\, k_x-\frac{\partial_x Q_{xx}(z,x)}{4 Q_{xx}(z,x)}+\frac{\partial_x Q_{zz}(z,x)}{4 Q_{zz}(z,x)}\,.
\end{split}
\end{equation}
The equations of motion for the momentum modes
\begin{equation}\label{kmodes}
\begin{split}
&\mathcal{F}_\alpha^{(n)}(z, \omega, k_x, k_y)= \left[ \begin{matrix} \mathcal{A}_{\alpha}^{(n)}(z,\omega,k_x,k_y) \cr  \mathcal{B}_{\alpha}^{(n)}(z,\omega,k_x,k_y) \end{matrix}\right],\quad \alpha=1, 2\,,\\
\mathcal{A}_{\alpha}=&\sum_{n=0,\pm1,\pm2} \mathcal{A}_{\alpha}^{(n)}\, e^{i n K x},\quad \mathcal{B}_{\alpha}=\sum_{n=0,\pm1,\pm2} \mathcal{B}_{\alpha}^{(n)}\, e^{i n K x},
\end{split}
\end{equation}
can be easily obtained after substituting~\eqref{kmodes} into~\eqref{Dirac1} and~\eqref{Dirac2}.

With the background geometry~\eqref{ansatzbh}, we find that the IR expansion near $z=0$ is of the form
\begin{equation}\label{irexpand}
\left[ \begin{matrix} \mathcal{A}_{\alpha}(z,x) \cr\mathcal{B}_{\alpha}(z,x) \end{matrix}\right]=z^{\pm\frac{i\omega}{2\pi T}}\left(
\left[\begin{matrix} \textbf{a}_{\alpha}^0(x)  \cr \textbf{b}_{\alpha}^0(x)\end{matrix}\right]+\left[\begin{matrix} \textbf{a}_{\alpha}^1(x)  \cr \textbf{b}_{\alpha}^1(x)\end{matrix}\right] z+\left[\begin{matrix} \textbf{a}_{\alpha}^2(x)  \cr \textbf{b}_{\alpha}^2(x)\end{matrix}\right] z^2+\cdots\right)\,,
\end{equation} 
with the minus sign choice corresponding to in-falling boundary conditions as required for the holographic computation of the retarded Green's function of the boundary theory. 
Note that the second term in parentheses is required by the nature of the singular points of the Dirac equations~\eqref{Dirac1} and~\eqref{Dirac2}. 
This is quite different from the IR expansion used to solve for the background functions~\eqref{ansatzbh}, where only even powers are needed in order to have a smooth horizon. 
Indeed, an expansion containing only even powers in the spinor functions $(\mathcal{A}_{\alpha}, \mathcal{B}_{\alpha})$ would result in an inconsistency, and thus has no solution. Two important relations for the expansion coefficients are obtained by substituting~\eqref{irexpand} into \eqref{Dirac1} and~\eqref{Dirac2},
\begin{equation}
\textbf{b}_{\alpha}^0(x)=-i \,\textbf{a}_{\alpha}^0(x),\quad \textbf{b}_{\alpha}^1(x)=i \,\textbf{a}_{\alpha}^1(x)\,,
\end{equation} 
which are used as the IR boundary condition when solving the Dirac equations numerically. 
In practice, we find that the second relation is important to avoid a badly conditioned matrix which could result in significant numerical errors.

On the other hand, near the AdS boundary $z=1$ the two Dirac equations~\eqref{Dirac1} and~\eqref{Dirac2}
reduce to
\begin{eqnarray} \label{uvexpand}
\partial_{z}\left[ \begin{matrix} \mathcal{A}_{\alpha}(z,x) \cr  \mathcal{B}_{\alpha}(z,x) \end{matrix}\right]
\pm \frac{mL}{1-z}
\left[ \begin{matrix} \mathcal{A}_{\alpha}(z,x) \cr  \mathcal{B}_{\alpha}(z,x) \end{matrix}\right]
=0\,,
\end{eqnarray}
and equivalently
\begin{eqnarray} \label{uvexpandn}
\partial_{z}\left[ \begin{matrix} \mathcal{A}_{\alpha}^{(n)}(z) \cr  \mathcal{B}_{\alpha}^{(n)}(z) \end{matrix}\right]
\pm \frac{mL}{1-z}
\left[ \begin{matrix} \mathcal{A}_{\alpha}^{(n)}(z) \cr  \mathcal{B}_{\alpha}^{(n)}(z) \end{matrix}\right]
=0\,.
\end{eqnarray}
We then obtain the following asymptotic expansion near the AdS boundary,
\begin{eqnarray} \label{boundary}
\mathcal{F}_\alpha=\left[ \begin{matrix} \mathcal{A}_{\alpha}(z,x) \cr  \mathcal{B}_{\alpha}(z,x) \end{matrix}\right]
 {=}  \; a_{\alpha}(x)(1-z)^{+mL}\left[ \begin{matrix} 1 \cr  0 \end{matrix}\right]
+b_{\alpha}(x)(1-z)^{-mL}\left[ \begin{matrix} 0 \cr 1 \end{matrix}\right]+\cdots\,,
\end{eqnarray}
and in terms of the momentum level
\begin{eqnarray} \label{boundaryn}
\mathcal{F}_{\alpha}^{(n)}=\left[\begin{matrix} \mathcal{A}_{\alpha}^{(n)} \cr  \mathcal{B}_{\alpha}^{(n)} \end{matrix}\right]
 {=} \; a_{\alpha}^{(n)}(1-z)^{+mL}\left[ \begin{matrix} 1 \cr  0 \end{matrix}\right]
+b_{\alpha}^{(n)}(1-z)^{-mL}\left[\begin{matrix} 0 \cr 1 \end{matrix}\right]+\cdots\,.
\end{eqnarray}
with $(a_{\alpha}^{(n)}, b_{\alpha}^{(n)})$ constants for a given $(\omega, k_x, k_y)$.
Finally, the retarded Green's function can be extracted by the following relation~\cite{Liu:2012tr,Ling:2013aya}\,\footnote{Notice that as we take $m\rightarrow-m$, we simply exchange the role of $(a_{\alpha}^{(n)}, b_{\alpha}^{(n)})$. We can thus restrict our attention to $m\geqslant 0$, for which $b_{\alpha}^{(n)}$ is identified as the source, while $a_{\alpha}^{(n)}$ as the response. The fermion operator in the dual field theory has scaling dimension $\Delta=\frac{3}{2}+mL$.}
\begin{equation}
a_{\alpha}^{(n)}(\omega, k_x, k_y)=\sum_{\alpha',n'}G^R_{\alpha,n;\alpha',n'}(\omega, k_x, k_y)\,b_{\alpha'}^{(n')}(\omega, k_x, k_y)\,.
\end{equation}

Note that in the spatially modulated case with periodic structure, the Green's function is characterized by two Bloch indices $(n, n')$, indicating contributions from different momentum levels or Brillouin zones.  Recall that in ARPES experiments the photoelectron propagates in the Galilean continuum and has a \emph{definite momentum}. Thus, we consider the Green's function in the momentum-basis. 
In the previous holographic studies~\cite{Liu:2012tr,Ling:2013aya} it was assumed that the main features of the spectral function are captured by the diagonal components of the retarded Green's function (although this now contains a mixing with other momentum modes). We have checked explicitly that the non-diagonal components are indeed quite small when the spatial modulation is weak. However, as the strength of the spatial modulation is increased, the non-diagonal components also increase\,\footnote{We find that at particular values of the momentum some of the off-diagonal components become comparable to the diagonal ones.}. This is reasonable since the non-diagonal components capture the interband interaction 
which is expected to be strong with a large spatially modulated potential.
Thus, there is a valid concern that by working only with the diagonal components of the Green's function one is neglecting the interaction 
between different Brillouin zones.
Formally, we could diagonalize the system into a new basis of modes. To do so explicitly is quite non-trivial because the system contains the full range of Bloch indices $n=0, \pm 1, \pm2,...$. 
However, we note that the trace (the sum of the diagonal components) remains invariant under unitary transformations (in particular diagonalization) 
and implicitly contains the effects associated with the non-diagonal components in the original basis. Therefore, in the present paper we consider the diagonal momentum spectral weight defined in terms of the trace and given by\,\footnote{For fixed $n,n'$, $G^R_{\alpha,n;\alpha',n'}$ is a $2\times2$ matrix in spin space, and each component of the matrix depends on the choice of representation (or gamma matrices). Since different representations are related by a unitary transformation, we consider the trace of the Green's function, which is invariant under the latter.}
\begin{equation}\label{spectral}
A(\omega, k_x,k_y)=\sum_{n=0,\pm1,\pm2,\cdots}\text{Tr}\; \text{Im}[G^R_{\alpha,n;\alpha',n}(\omega, k_x,k_y)]\,.
\end{equation}
Here $k_x\in[-\frac{K}{2},\frac{K}{2}]$ and $n$ denotes once again the momentum level or Brillouin zone. Note that the spectral density (also known as spectral function) $A(\omega, k_x,k_y)$ should be positive as required by unitarity. 

Ideally, we would like to study fermions associated with consistent truncations of UV-complete theories. However, for lack of a better construction we will content ourselves with a bottom-up approach in which there is no particularly good reason to choose specific values of the charge and mass of the bulk fermion. 
We will take $m=0$ and consider different values of the charge, which corresponds to  
scanning through different dual boundary field theories. 
As we will see in the next section, the behavior of the femionic spectral density depends significantly on some of the
theory parameters. Nevertheless, it is still possible to identify interesting properties that appear to be generic.

%%%%%%%%%%%%%%%%%%%%%%%%%%%%%
%%%%%%%%%%%%%%%%%%%%%%%%%%%%%
\section{Numerical Results}
\label{Numerics}
%%%%%%%%%%%%%%%%%%%%%%%%%%%%%
%%%%%%%%%%%%%%%%%%%%%%%%%%%%%

The location of the Fermi surface is typically identified as a pole in the spectral density at zero temperature as $\omega\rightarrow 0$. 
Accessing numerically the ground state $T=0$ geometry in our setup is unfortunately very challenging.
Thus, we will work instead at finite but low temperature.
Although in this case one can not expect a true  Fermi surface singularity in the spectral density, the presence of the Fermi surface should still be indicated by a (sufficiently strong) peak in the spectral density. 
In particular, to judge whether a holographic Fermi surface exists or not at finite temperature, we will apply the width, frequency and magnitude criteria introduced in~\cite{Cosnier-Horeau:2014qya}. 

The width criterion demands that the width of the peak in the spectral density at $\omega\rightarrow 0$ should be no greater than an $\mathcal{O}(1)$ factor times the temperature.  
As a consequence, a peak that is very broad compared to $T$ will not be regarded as evidence of a Fermi surface. The frequency criterion states instead that if a maximum at $\vec{k}=\vec{k}_*$ is to be regarded as a Fermi surface, then the spectral density as a function of $\omega$ should show a peak with a maximum near $\omega =0$. 
This is consistent with the presence of a quasi-particle near the Fermi surface. The last one is a more heuristic criterion: the magnitude of the spectral density should be ``large'' at $\vec{k}=\vec{k}_*$ as $\omega\rightarrow 0$.

In this section we will study the behavior of the spectral density and identify the presence of a Fermi surface by applying the criteria stated above. 
We will examine three different cases: 
\begin{itemize}
\item
Case (i): the spatial modulations that break translational invariance are generated spontaneously without any source; 
\item
Case (ii): an ionic lattice is introduced explicitly via~\eqref{lattice} on top of the spontaneously generated striped background.
\item
Case (iii): an ionic lattice is added explicitly in the standard Einstein-Maxwell theory without spontaneously generated stripes.
\end{itemize}
We shall fix $m=0$ for numerical convenience. 
As can be seen from figures~\ref{fig:geometry} and \ref{fig:geometryIonic}, the corresponding configurations of the metric components and gauge field $\alpha$ have a period of $\pi/k$ along the symmetry broken direction. Since these are background functions that enter into the bulk Dirac equations~\eqref{Dirac1} and~\eqref{Dirac2}, the Umklapp wavevector $K$ felt by the probe fermion is $K=2k$. 
We will identify the location of the Fermi surface by searching for the peaks of $A(\omega, k_x, k_y)$ satisfying the width, frequency and the magnitude criteria and working at frequencies $\omega$ that are very close to zero.

\subsection{Case (i): PDW without Ionic Lattice}

We start by discussing properties of the spectral density for the pure PDW phase without an ionic lattice. In this case the spatial modulations are generated spontaneously and the geometry 
is shown in figure~\ref{fig:geometry}.

\subsubsection{Charge dependence of the momentum distribution}
\begin{figure}[ht!]
\begin{center}
\includegraphics[width=.50\textwidth]{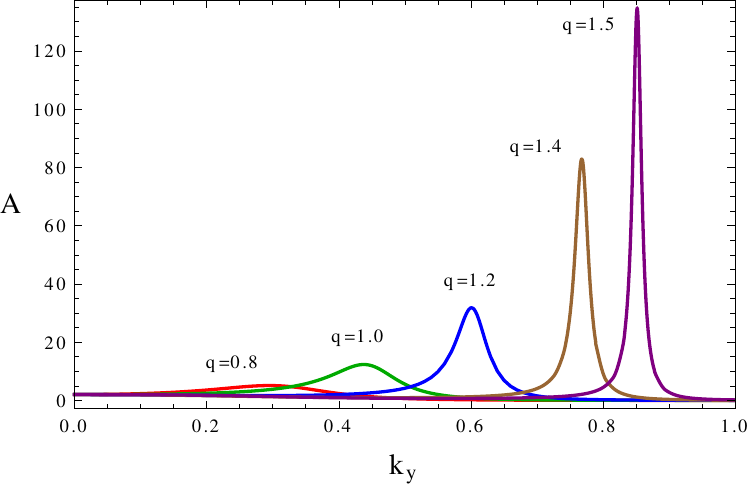}\\
\includegraphics[width=.46\textwidth]{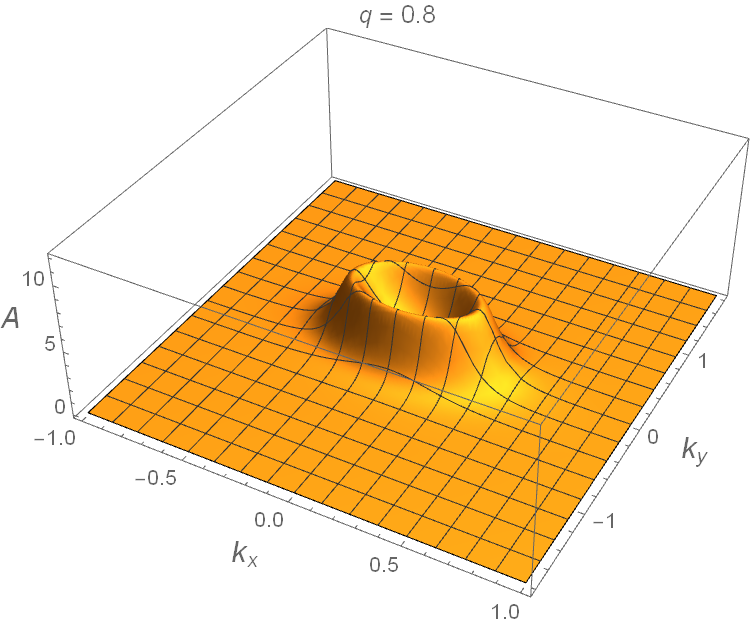}\qquad
\includegraphics[width=.46\textwidth]{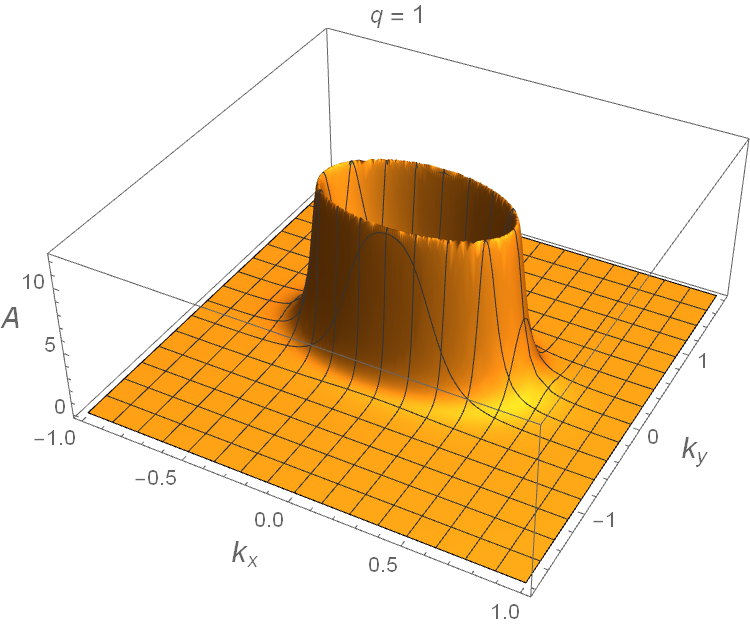}
\caption{Top: Momentum distribution of the spectral density (MDF) as a function of $k_y$ for fixed $k_x=0$ and varying values of $q$. Bottom: The 3D plots of the spectral density in momentum space $(k_x, k_y)$ for $q=0.8$ and $q=1$. 
We choose $\omega=10^{-6}$ and work with the background geometry shown in figure~\ref{fig:geometry}.}
\label{fig:specvsky}
\end{center}
\end{figure}

\begin{figure}[ht!]
\begin{center}
\includegraphics[width=.45\textwidth]{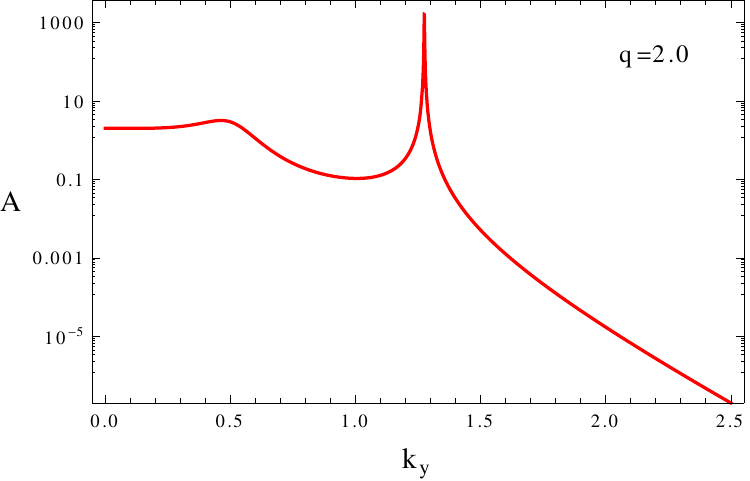}\quad\quad
\includegraphics[width=.45\textwidth]{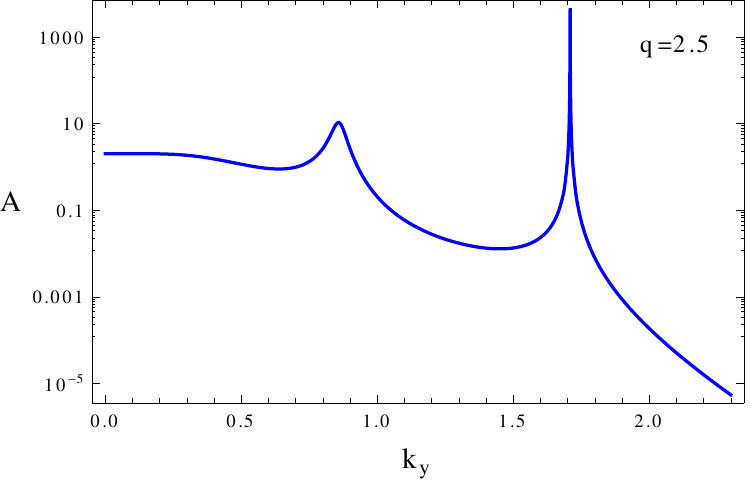}
\caption{MDF spectral density as a function of $k_y$ for $q=2.0$ (left) and $q=2.5$ (right).  We have fixed $k_x=0$ and $\omega=10^{-6}$. 
In each plot there is a very sharp spectral weight indicative of a Fermi surface. Note that the vertical axis is logarithmic, 
causing the peak to have a spike-like appearance. 
}
\label{fig:specvsky2}
\end{center}
\end{figure}

The momentum distribution function (MDF), i.e., the spectral density as a function of momentum, is plotted in figures~\ref{fig:specvsky} and~\ref{fig:specvsky2} 
for different values of the charge $q$ of the bulk fermion. 
We see that the spectral density develops a peak whose amplitude increases as $q$ is increased. 
When $q$ is small, the peaks are very broad compared to $T$ and therefore should not be regarded as evidence for a Fermi surface, according to the criteria we discussed at the beginning of this section.
On the other hand, when $q$ is sufficiently large the peaks are very sharp and satisfy all criteria, as can be seen from figure~\ref{fig:specvsky2} -- 
we conclude that there is a Fermi surface.
In figure~\ref{fig:specvsky2}  we also see that for large $q$ the spectral density tends to develop a more complicated structure, 
with the appearance of additional small peaks -- an indication
that additional Fermi surfaces will likely form for sufficiently large charge.\footnote{An analytic formula for Fermi momenta was found for a specific system~\cite{Gubser:2012yb} as $k_F^{(n)}=q-n-1/2$, where $n$ is a nonnegative integer. It clearly shows the absence of Fermi surface for small $q$ and the appearance of multiple Fermi surfaces for large $q$. This is qualitatively in agreement with the AdS-RN black hole~\cite{Herzog:2012kx}.}

\subsubsection{Fermi surface and band gap}
\begin{figure}[ht!]
\begin{center}
\includegraphics[width=.55\textwidth]{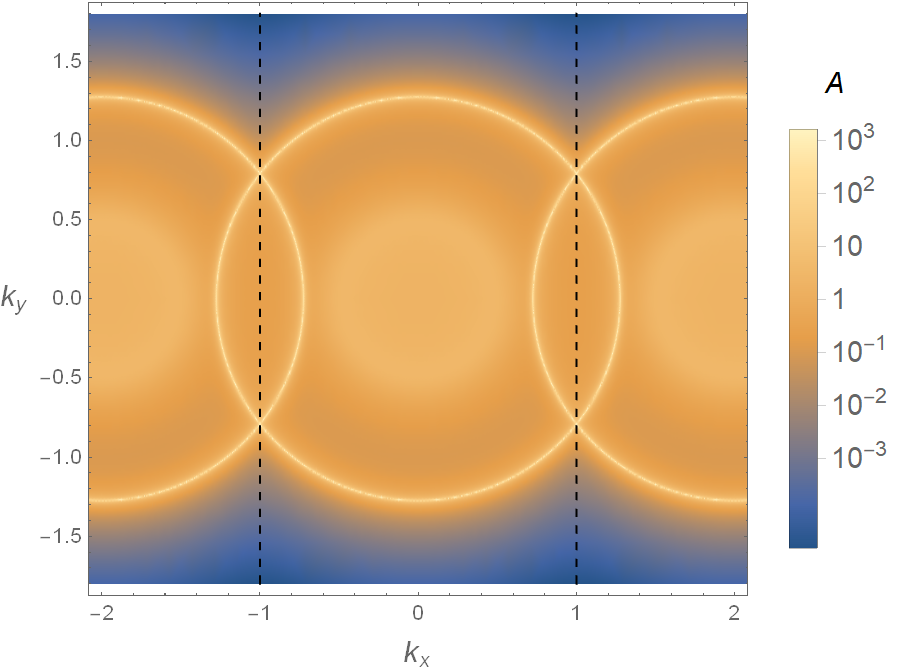}
\caption{The density plot of the MDF in the $(k_x, k_y)$ plane for $q=2$ and $\omega=10^{-6}$. The brightest points correspond to the location of the Fermi surface. We have used a logarithmic scale and periodically extended the data from the first Brillouin zone to the other ones.
The first Brillouin zone boundary is denoted by the vertical dashed lines at $k_x=\pm 1$, and the background geometry has $T=0.01426$ and $k=1$.}
\label{fig:densityq2}
\end{center}
\end{figure}

Our main interest in this paper is in the formation and structure of the Fermi surface, including the presence of a possible band gap, in striped superconducting phases. 
For concreteness in the rest of the discussion we will focus on $q=2.0$, a value for the charge large enough to support a Fermi surface. 
The density plot of the corresponding MDF is shown in figure~\ref{fig:densityq2}. 
We emphasize that we compute the spectral density $A(\omega, k_x, k_y)$ in the first Brillouin zone $k_x\in[-\frac{K}{2},\frac{K}{2}]$, and simply periodically extend the result to the other Brillouin zones in figure~\ref{fig:densityq2}.
Note that as the strength of the PDW modulation increases, the shape of the Fermi surface will become more anisotropic.

\begin{figure}[ht!]
\begin{center}
\includegraphics[width=.50\textwidth]{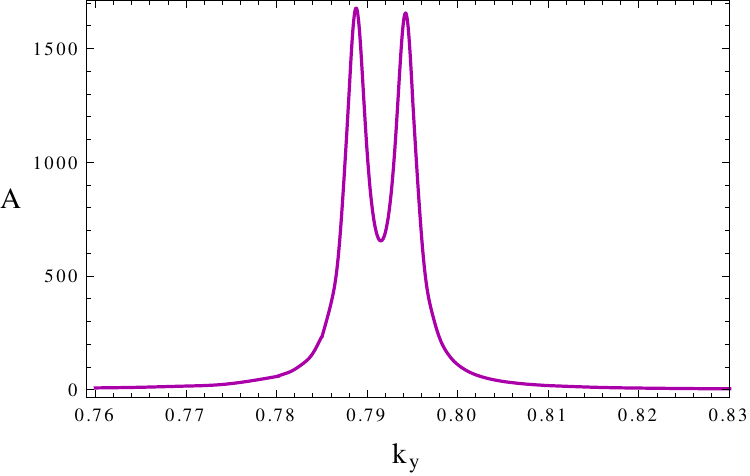}
\caption{The spectral density $A$ as a function of $k_y$ at the first Brillouin zone boundary $k_x=1$ for $q=2, \omega=10^{-6}$. There are two pronounced peaks near $k_y=0.79$ 
 indicative of two separate Fermi surface  branches. The background geometry is that of figure~\ref{fig:geometry} with $T=0.01426$ and $k=1$.}
\label{fig:bandgap}
\end{center}
\end{figure}

It is well known that, for degenerate eigenvalues at the Brillouin boundary, a band gap opens up due to eigenvalue repulsion. Therefore, when the Fermi surface intersects the first Brillouin zone at $k_x=\pm 1$,  
one anticipates to see a similar gap structure  in the behavior of the spectral density, due to the broken translational invariance and the periodic modulation of the background.
Indeed, once we zoom in near the Fermi surface at the 
Brillouin boundary and inspect the spectral density, shown in figure~\ref{fig:bandgap}, we find two sharp peaks. 
These indicate two Fermi surface
branches in figure~\ref{fig:densityq2}, an inner and an outer one, with a gap between them which is minimal at the 
Brillouin zone boundary. 
In addition to the sharp peaks, in figure~\ref{fig:densityq2} we also see a small circle of broad peaks, which however fail to satisfy the criteria we discussed above
and thus should not be identified with Fermi surfaces. 
We anticipate from the behavior seen in figure~\ref{fig:specvsky2} that such bumps will become sharp as $q$ is increased, and develop into additional Fermi surfaces when $q$ is sufficiently large.

\begin{figure}[ht!]
\begin{center}
\includegraphics[width=.60\textwidth]{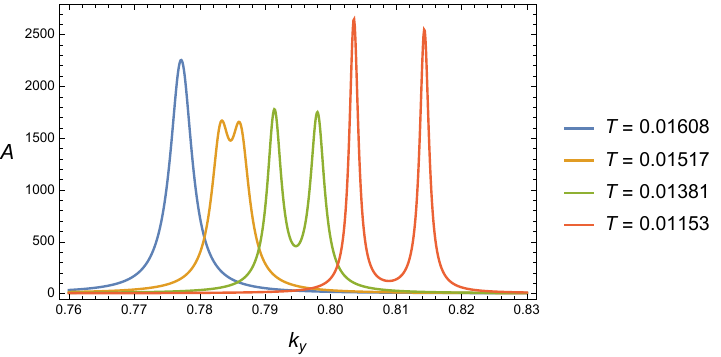}
\caption{Evolution of the band gap with temperature for the pure PDW case. We show the spectral density as a function of $k_y$ at the Brillouin boundary for fixed $\omega=10^{-6}$. From left to right, the four curves correspond to temperatures $T/T_c=1$, $0.94$, $0.86$, and $0.72$.}
\label{fig:gapvsT}
\end{center}
\end{figure}

Recall that in our construction the spatially modulated background is generated spontaneously below the critical temperature $T_c = 0.016$. 
A natural question to ask, then, is what happens to the gap as the temperature decreases. 
In particular, since the amplitude of the modulation increases as the temperature is lowered, we expect that the gap should develop at $T_c$ and become large at low temperatures. To verify this, we examine the temperature dependence of the band gap in figure~\ref{fig:gapvsT}. Just as expected, at $T_c$ the spectral density has only one peak. Below $T_c$  the peak begins to split into two and the separation between them grows up as $T$ is decreased, confirming our intuition.

For a discussion of the energy distribution of the spectral density we refer the reader to Appendix \ref{appA}.

\subsection{Case (ii): PDW with Ionic Lattice}

Next, we consider the case in which a periodic background potential breaks translational invariance explicitly. To do so, we introduce an ionic lattice 
through a spatially modulated chemical potential~\eqref{lattice} in the field theory (see figure~\ref{fig:geometryIonic} for representative bulk profiles).
We are particularly interested in highlighting the features that differ from those of the pure PDW case. The issue of how strong lattice potentials influence the fermion
spectral functions within the framework of holography is still an open question. Moreover, it provides a crucial test of the applications of holographic techniques to condensed matter
materials. Below we will show the behavior of the spectral function as the strength of the translational symmetry breaking potential is increased, 
and identify interesting features.

\subsubsection{Charge dependence of the momentum distribution}

\begin{figure}[ht!]
\begin{center}
\includegraphics[width=.50\textwidth]{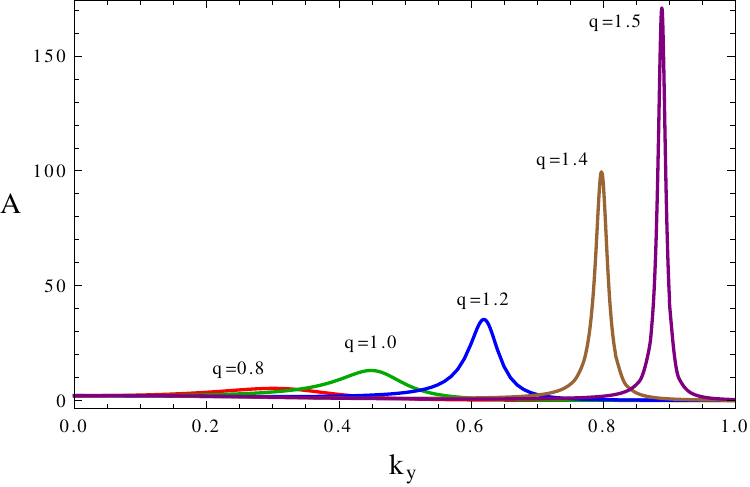}\\
\includegraphics[width=.45\textwidth]{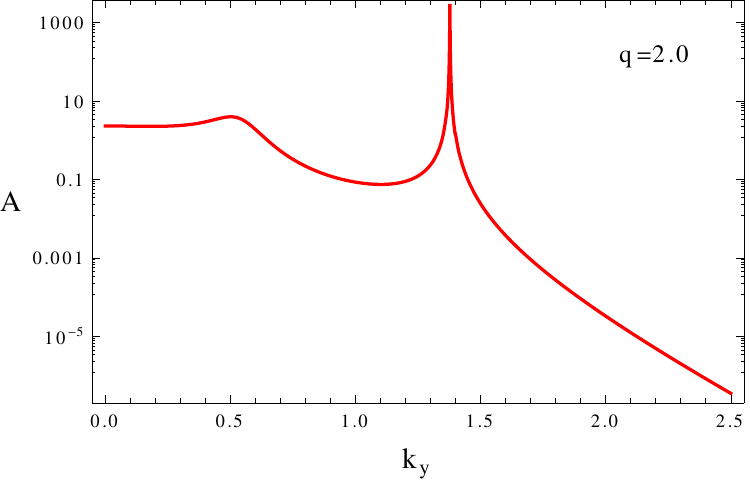}\quad\quad
\includegraphics[width=.45\textwidth]{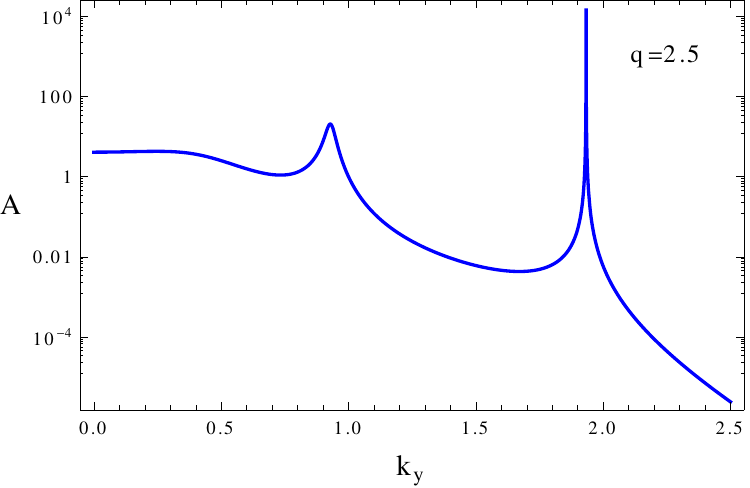}
\caption{The MDF as a function of $k_y$ at $(k_x=0, \omega=10^{-6})$ for different values of the charge $q$ of the fermion. Note that in the last two plots the vertical axis is logarithmic, and there is a very sharp peak indicative of a Fermi surface. The background geometry is the same as in figure~\ref{fig:geometryIonic} with the ionic lattice amplitude $a_0=0.5$ and the wavenumber $p=2$.}
\label{fig:specvskyionic}
\end{center}
\end{figure}

The spectral density $A$ as a function of $k_y$ is presented in figure~\ref{fig:specvskyionic} for different values of the charge $q$ of the bulk fermion. 
The behavior is similar to that of the pure PDW phase. 
For each charge, the spectral density develops a peak at a certain value of $k_y$, whose amplitude increases as $q$ is increased. Once again, there are Fermi surfaces when $q$ is sufficiently large. 
Note that compared to the pure PDW case, the amplitude of the spectral density along the $k_y$ axis is enhanced after turning on the ionic lattice. Thus, the explicit breaking seems to slightly facilitate the formation of a Fermi surface.

\subsubsection{Fermi surface, band gap and gradual disappearance}

The sharpest difference between the pure PDW phase and the case with an explicit lattice 
comes into play when we examine the behavior of the Fermi surface as the strength of the 
ionic lattice is varied. As we will see below, the band gap grows with the amplitude of the lattice. 
Moreover, increasing the strength of the latter also causes a gradual disappearance of the Fermi surface along the symmetry breaking direction, eventually leading to the formation of small disconnected Fermi surface segments. 
These features are quite distinct from the pure PDW case without explicit  sources of symmetry breaking.
However, this difference may simply be due to the fact that the magnitude of the PDW modulations is significantly smaller than that of the UV lattice, at the 
temperatures we work with. Repeating the analysis at much lower temperatures would clarify the origin of this effect, and in particular whether it is generically associated with broken translational 
invariance, independently of whether it is spontaneous or explicit.

\begin{figure}[ht!]
\begin{center}
\includegraphics[width=.55\textwidth]{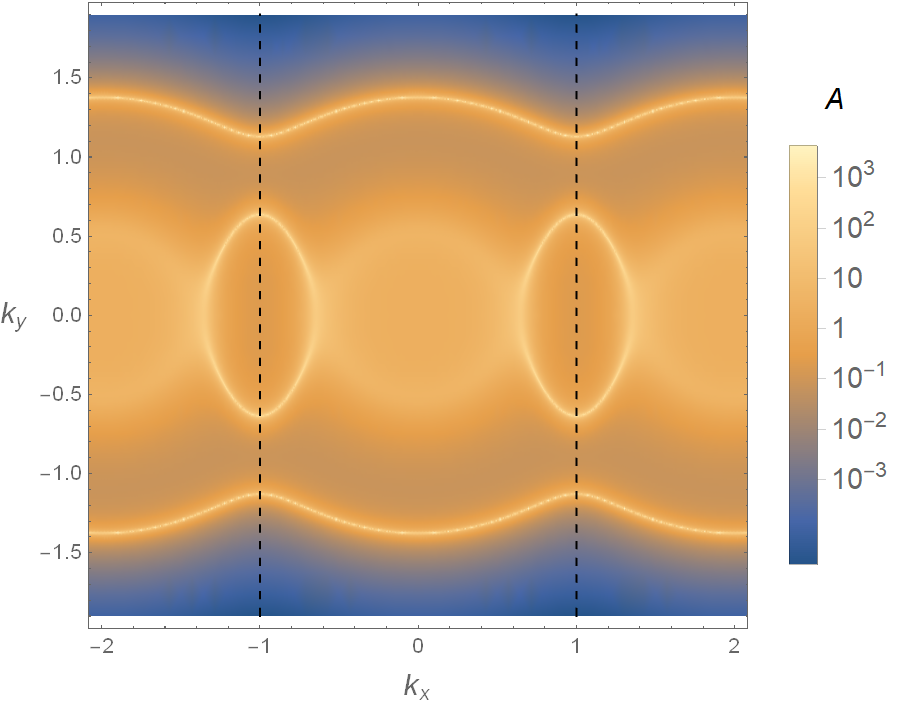}
\caption{The density plot of $A$ in the $(k_x, k_y)$ plane after turning on the ionic lattice for fixed $\omega=10^{-6}$, $q=2$. The brightest points correspond to the location of the Fermi surface. We have adopted a logarithmic scale and periodically extended the data from the first Brillouin zone to the other zones. 
This case corresponds to the PDW phase shown in figure~\ref{fig:geometryIonic} with $T=0.01426$, $a_0=0.5$, $p=2$. Note that the vertical axis is logarithmic. The first Brillouin zone boundary is indicated by two vertical dashed lines located at $k_x=\pm 1$.}
\label{fig:densityq2ionic}
\end{center}
\end{figure}
The density plot of the MDF is shown in figure~\ref{fig:densityq2ionic} for the PDW phase in the 
presence of an ionic lattice. We see that once again the Fermi surface consists of two branches, as in the case without explicit symmetry breaking. 
Compared to figure~\ref{fig:densityq2}, we find a much more pronounced band gap at the Brillouin zone boundary.

\begin{figure}[ht!]
\begin{center}
\includegraphics[width=.60\textwidth]{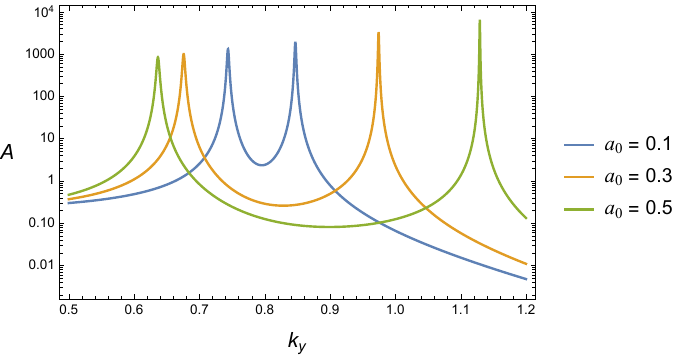}
\caption{The band gap with respect to the amplitude of the ionic lattice. The larger the lattice potential, the larger the gap that is observed. We choose $\omega=10^{-6}$, $q=2$ as well as the background geometry with $T=0.01426$, $p=2$. Note that the spike-like appearance of the peaks is due to the logarithmic scale.}
\label{fig:gapvsa0}
\end{center}
\end{figure}

In particular, the size of the band gap as a function of the amplitude 
of the ionic lattice is shown in figure~\ref{fig:gapvsa0}. Another feature one can see in figure~\ref{fig:gapvsa0} is that the peak associated with the outer Fermi surface is enhanced, while the inner one is slightly reduced -- we don't 
know whether this is a generic effect or if it is model dependent.

\begin{figure}[ht!]
\begin{center}
\includegraphics[width=.65\textwidth]{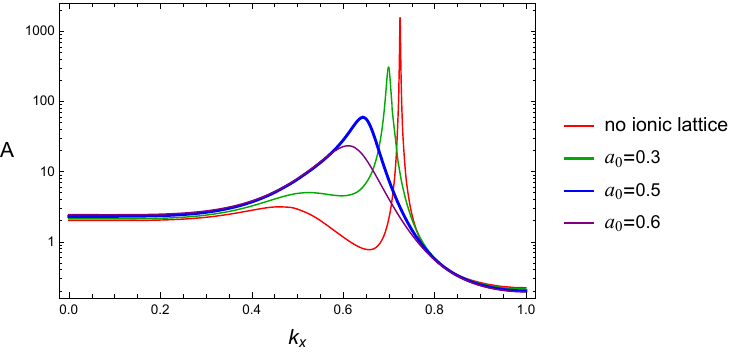}
\caption{The MDF along the $k_x$-axis (when $k_y=0$) for different values of the amplitude of the ionic lattice  for fixed $\omega=10^{-6}$, $q=2$ and $T=0.01426$. The case with $a_0=0.5$ is denoted by a thick blue curve. The first Brillouin zone boundary is at $k_x=1$. This shows that the inner Fermi surface dissolves as the strength of the ionic lattice increases.}
\label{fig:AxskxIonic}
\end{center}
\end{figure}

Moreover, as we examine carefully the density plot in figure~\ref{fig:densityq2ionic} along the horizontal axis, we find that the inner Fermi surface seems to be partially dissolved, i.e. 
the peak of the MDF becomes smooth and broad. 
To make this effect more visible, in figure~\ref{fig:AxskxIonic} we show the behavior of the MDF along the symmetry breaking direction (i.e. as a function of $k_x$) as the amplitude of the ionic lattice becomes larger (we set $k_y=0$).
It is clear that the peak of the spectral weight becomes weaker and broader as the strength of the  lattice is increased. Therefore, for sufficiently large lattice amplitude we find that the \emph{inner} Fermi surface is no longer closed,  but rather appears to consist of detached segments, a behavior 
which is reminiscent of Fermi arcs~\cite{Norman,Kanigel1,Kanigel2}.

\begin{figure}[ht!]
\begin{center}
\includegraphics[width=.66\textwidth]{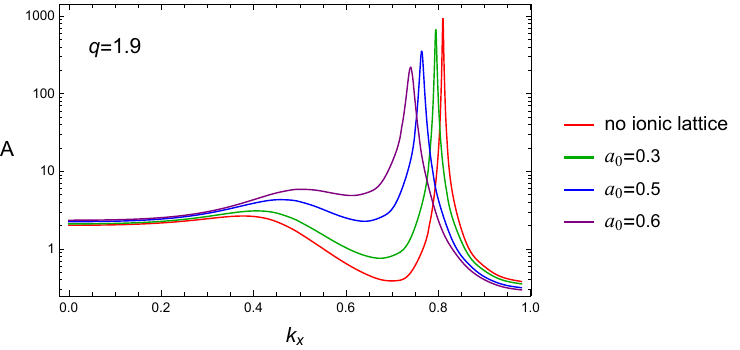}
\caption{The MDF along the $k_x$-axis for different values of the amplitude of the ionic lattice for $q=1.9$. The first Brillouin zone boundary is at $k_x=1$. The inner Fermi surface dissolves as the strength of the ionic lattice increases.}
\label{fig:AxskxIonicq19}
\end{center}
\end{figure}

Intriguingly, figure~\ref{fig:AxskxIonic} clearly shows that the peak of the ``inner Fermi surface" gets broad at the same time as it merges with the broad
feature due to the secondary Fermi surface\,\footnote{The broader central peak would eventually develop into a Fermi surface if the charge of the fermion 
was increased sufficiently. However, at this particular fixed charge it is \emph{not} a Fermi surface.}. Thus, this raises the question of whether the broadening effect that leads to the gradual disappearance of the Fermi surface is tied \emph{generically} to the existence of a secondary surface and the merging of the two peaks.
To examine this point further and check the relevance of the secondary Fermi surface on the ``Fermi arc'' effect, we push the leading and the secondary Fermi surfaces apart by choosing a smaller value of $q$. The case with $q=1.9$ is presented in figure~\ref{fig:AxskxIonicq19}, where it is clear that the two peaks are separated by a larger distance. Once again we find that the Fermi surface gradually disappears as the strength of the ionic lattice is increased. Meanwhile, the two peaks associated with the Fermi surface and the secondary surface move towards each other. However, compared to the case with $q=2.0$ a much stronger lattice is needed to merge them. 
Moreover, we note that in the example studied in subsection~\ref{appC} the disappearance of the Fermi surface does \emph{not} appear to be tied to the merging of the peaks (see figure~\ref{fig:specvskxmaxwell}).  
Thus, from the charge dependence seen in figures~\ref{fig:AxskxIonic}  and~\ref{fig:AxskxIonicq19} and more importantly from the behavior in figure~\ref{fig:specvskxmaxwell} we are led to conclude that this merging phenomenon is \emph{not} generically responsible for the Fermi arc effect. Further work is needed to fully clarify the relevance of this phenomenon and to reach a more detailed understanding of the role of the secondary peak.

\begin{figure}[ht!]
\begin{center}
\includegraphics[width=.65\textwidth]{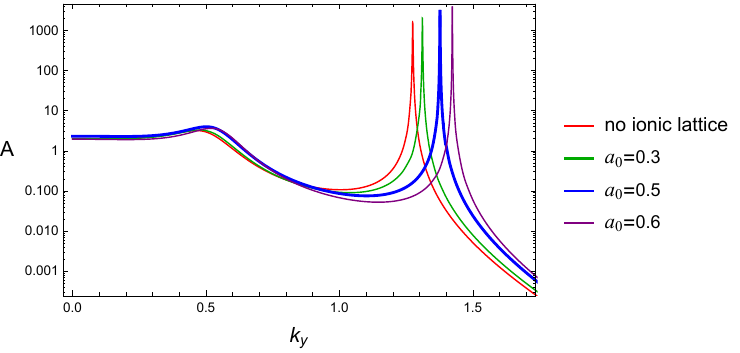}
\caption{The MDF along the $k_y$-axis for varying values of the amplitude of the ionic lattice  for fixed $\omega=10^{-6}$, $q=2$ and $T=0.01426$. The case with $a_0=0.5$ is denoted by a thick blue curve. The first Brillouin zone boundary is at $k_x=1$.}
\label{fig:AxskyIonic}
\end{center}
\end{figure}

In contrast to what happens to the MDF along the symmetry breaking direction $k_x$, 
the behavior along $k_y$ is not affected by the explicit lattice, and the \emph{outer} Fermi surface is still present
as the amplitude of the ionic lattice is increased. As shown in figure~\ref{fig:AxskyIonic}, the spectral weight along $k_y$ (the direction which respects translational invariance) is enhanced for large lattice amplitudes. 

One might wonder whether it is the interplay between the PDW and the ionic lattice that gives rise to the ``Fermi arcs." 
To check whether this is true, we turn next to a special case of our theory which describes an explicit ionic lattice but without any spontaneous symmetry breaking (no PDW).
As we will see, even in that case we find a gradual disappearance of the Fermi surface as the strength of the lattice increases.

%%%%
%

\subsection{Case (iii): Fermionic Spectral Function with Only an Ionic Lattice}
\label{appC}

As we have shown above, when only spontaneous order is considered (the pure PDW case) 
at the temperatures we have studied the resulting Fermi surface does not display any ``Fermi arcs". 
On the other hand, when one includes the ionic lattice in the UV and takes its amplitude to be large enough, the spectral function along the symmetry broken direction is suppressed, leading to the appearance of ``Fermi arcs" (see figure~\ref{fig:densityq2ionic}). 
We can draw three possible conclusions from these two examples. 
The first one is that the spontaneous PDW order plays no role in this process, and that the ionic lattice 
is entirely responsible for the destruction of the Fermi surface. 
The second explanation is that the ``Fermi arcs" are in fact strictly due to the interplay between both PDW and ionic lattices, and would not occur if we only had  an explicit source of symmetry breaking.
Finally, it is also possible that both spontaneous and explicit breaking of translations play a role, but at the temperatures we have studied the role of the spontaneously generated modulation is simply not visible.

In this subsection we rule out the second possibility and show that the Fermi arc phenomenon seems to be a generic signature of strong (explicit) translational symmetry breaking. 
In particular, we consider an example with an ionic lattice without any PDW, and show that the segmentation of the Fermi surface is already visible there. 
However, it is important to emphasize that in order to distinguish between the first and third scenarios 
we would need to construct the background geometry at much lower temperatures, 
to ensure that the spontaneously generated modulation in the IR is strong enough to compete with the explicit UV modulation. Indeed, it is still possible that a 
sufficiently strong spontaneous modulation would lead to the suppression of the spectral weight.

By setting $Z_A=1$ and $Z_B=Z_{AB}=\mathcal{K}=V=0$ in ~\eqref{actions} and turning off the scalar $\chi$ we arrive at the standard Einstein-Maxwell theory,
\begin{eqnarray}
S=\frac{1}{2\kappa_N^2}\int d^{4}x \sqrt{-g} \left[\mathcal{R}+\frac{6}{L^2}-\frac{1}{4}F_{\mu\nu}F^{\mu\nu}\right].
\end{eqnarray}
Since this is just a special case of the model we have considered in this paper, we can still use our setup in Section~\ref{backreaction} and Section~\ref{Greenfunction} to construct the background geometry and to compute the spectral function. In the present case, however, there is no spontaneous order and the spatial modulation is introduced explicitly by adding the UV ionic lattice~\eqref{lattice}. As a typical example, we consider the ionic lattice
\begin{equation}
\mu(x)=A_t(1,x)=2.35 [1+a_0 \cos(4\, x)]\,,
\end{equation}
which means that $\mu=2.35$ and the Umklapp wavevector $K$ felt by the probe fermion is $K=4$ with the first Brillouin zone boundary at $k_x=\pm2$.

\begin{figure}[ht!]
\begin{center}
\includegraphics[width=.45\textwidth]{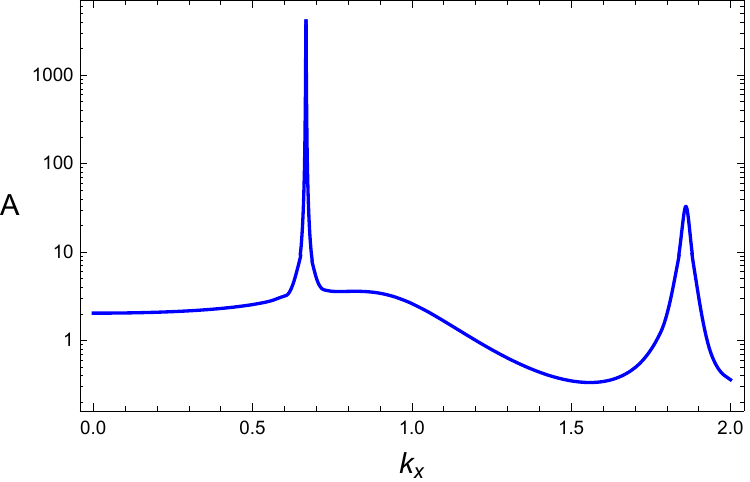}\quad\quad
\includegraphics[width=.45\textwidth]{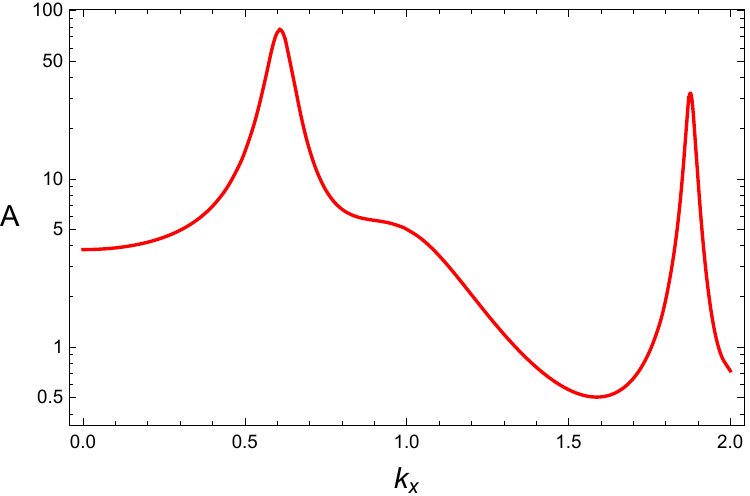}
\caption{The MDF along the $k_x$-axis for the amplitude of the ionic lattice $a_0=0.1$(left plot) and $a_0=0.8$ (right plot) for fixed $\omega=10^{-6}$, $q=2$ and $T=0.019$. The first Brillouin zone boundary is at $k_x=2$. This shows that the inner Fermi surface dissolves as the strength of the ionic lattice increases. Note that the vertical axis is logarithmic, and that there is a very sharp peak indicative of a Fermi surface. We have chosen $L=1$ and $\mu=2.35$.}
\label{fig:specvskxmaxwell}
\end{center}
\end{figure}

\begin{figure}[ht!]
\begin{center}
\includegraphics[width=.45\textwidth]{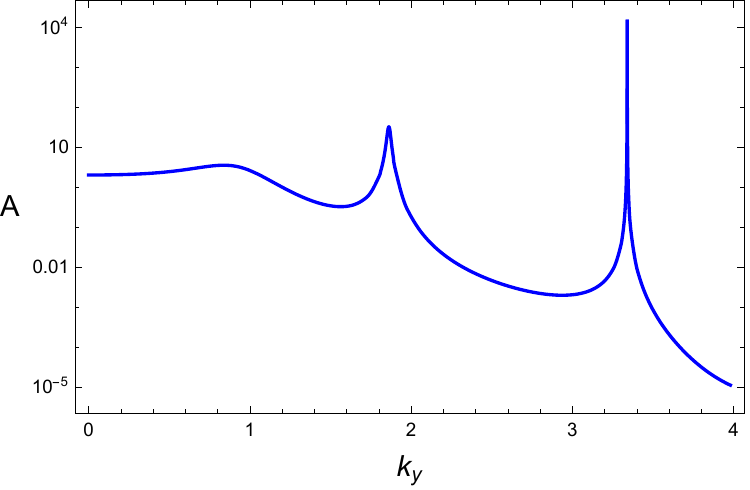}\quad\quad
\includegraphics[width=.45\textwidth]{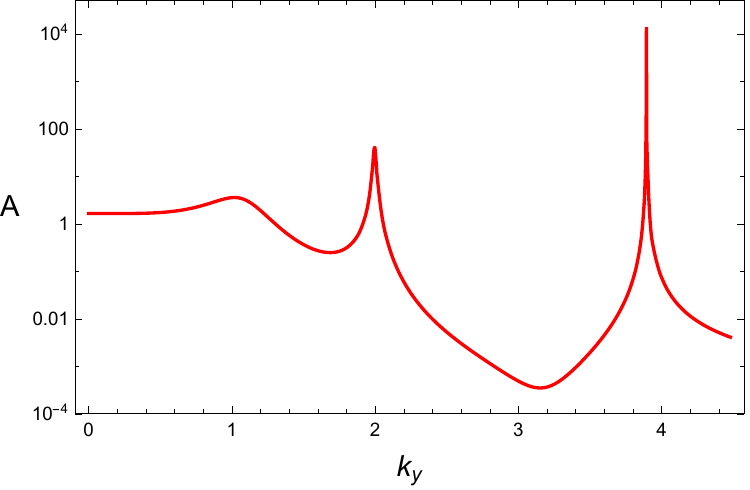}
\caption{The MDF along the $k_y$-axis for the amplitude of the ionic lattice $a_0=0.1$(left plot) and $a_0=0.8$ (right plot) for fixed $\omega=10^{-6}$, $q=2$ and $T=0.019$.  Note that the vertical axis is logarithmic, and that there is a very sharp peak indicative of a Fermi surface. We have chosen $L=1$ and $\mu=2.35$.}
\label{fig:specvskymaxwell}
\end{center}
\end{figure}

In figure~\ref{fig:specvskxmaxwell} we present the behavior of the spectral function along the symmetry breaking direction (i.e. as a function of $k_x$) as the amplitude of the ionic lattice becomes large. For small ionic amplitude (left plot) there is a very sharp peak indicating the appearance of a Fermi surface. However, as the strength of the lattice is increased the peak of the spectral weight becomes weaker and broader (right plot).  In contrast to the suppression of the spectral weight along the symmetry breaking direction $k_x$,  the behavior along $k_y$ is not affected by the explicit lattice and the Fermi surface is still present as the amplitude of the ionic lattice is increased, see figure~\ref{fig:specvskymaxwell}.
Therefore, for sufficiently large lattice amplitude we find that the Fermi surface is no longer closed,  but rather appears to consist of detached segments, which is reminiscent of Fermi arcs~\cite{Norman,Kanigel1,Kanigel2}. 
We emphasize that this behavior in this particular model is due entirely to the strength of the explicit UV lattice.

We can draw several lessons from this section.
For the cases we have considered thus far the spontaneous PDW order does not play a key role in the destruction of the Fermi surface. Nevertheless, it is still possible that a sufficiently strong spontaneous modulation at low temperature would lead to the suppression of the spectral weight.
We anticipate that in the fully 2D crystallized phase, in which translational invariance is broken along both spatial directions (see~\cite{Cai:2017qdz,Withers:2014sja,Donos:2015eew} for a full construction), we would obtain detached segments of the Fermi surface analogous to the Fermi arcs seen in PDW phases. We stress that the appearance of these disconnected arcs seems to be tied to the translational symmetry breaking mechanism. 
Indeed, our numerical results suggest that \emph{in strongly correlated systems Fermi surfaces can be suppressed when the inhomogeneity effect is strong enough}. 
To confirm this intuition in other settings it would be necessary to construct additional types of low temperature background geometries and study the associated fermionic response.
In the concluding section we will comment on the possible 
physical interpretation for the gradual disappearance of the Fermi surface along the symmetry breaking direction, at large lattice, and outline directions for examining this feature in further detail in future work.

\section{Conclusions}
\label{Conclusions}

This analysis is a preliminary step towards probing the Fermi response of a strongly coupled quantum system in which 
translational invariance is broken spontaneously (with and without an underlying ionic lattice). 
We have worked with a holographic model of a striped superconducting phase 
which shares certain key features of PDW order. 
In addition, we have added to the model a UV source which breaks translational invariance explicitly and describes an ionic lattice.
The final construction describes spontaneous crystallization in the presence of a background lattice, thus providing
a concrete framework to examine the interplay of 
spontaneous and explicit  breaking of translations on the fermionic response in the system.

We have identified several main features, some of which confirm previous results in the literature and therefore provide good checks on our analysis. 
The most physically interesting result we find, which is novel, is the disappearance of the Fermi surface with increasing 
lattice strength.
We summarize our results here, in the order in which they are discussed in the article (for an analysis of the
energy distribution of the spectral density see Appendix \ref{appA}):

\emph{(i) Charge dependence and Fermi surface formation:} the existence and size of the Fermi surface are both sensitive to the charge of the bulk fermion, 
as already seen in other contexts in the literature (the appearance of new Fermi surface branches as the charge is varied was already 
noticed in the early works on holographic non-Fermi liquids, starting with \cite{Liu:2009dm}).
For the cases we have studied, the Fermi surface can form and grow in size only once the charge $q$ is sufficiently large.
This is true independently of whether the system is in a pure PDW phase, or whether it contains an additional ionic lattice. 
However, in the latter case the 
amplitude of the spectral density is enhanced, compared to the pure PDW. Thus,  the explicit breaking seems to slightly facilitate the formation of a Fermi surface.
The charge dependence of the Fermi surface confirms our expectations from prior results in the literature.

\emph{(ii) Fermi Surface shape and band gap:}
when the Fermi surface is large enough to cross the Brillouin zone boundary, its shape is modified -- a gap develops at the zone boundary due to the periodic modulation of the background geometry (see figure~\ref{fig:densityq2} and figure~\ref{fig:densityq2ionic}). Note that this feature is not visible in the homogeneous case 
or in homogeneous lattices, as neither one can capture the physics of Umklapp. 
This is a basic feature characterizing the behavior of fermions in periodic potentials and was previously seen in~\cite{Ling:2013aya}. The fact that our results reproduce the
expected Umklapp gap is therefore a good check on our analysis.
In the pure PDW case without the explicit UV lattice, the gap increases as the temperature is lowered (and the amplitude of the spontaneous modulation increases).
The size of the gap also grows with the strength of the UV lattice. 
These behaviors are anticipated since either decreasing the temperature or increasing the amplitude of the ionic lattice results in a much larger periodic deformation of the background geometry  
the fermion lives in.
However, we should mention that in our analysis the increase in the gap is more apparent as the explicit symmetry breaking parameter grows.
We expect that this is simply due to the fact that we haven't reached temperatures low enough for the magnitude of the spontaneously generated background oscillations to compete with those of the 
UV lattice. This should change as lower temperatures are reached, and the IR and UV effects start being of comparable size.

\emph{(iii) Gradual disappearance of the Fermi surface:}
a more intriguing feature emerging from our analysis is the suppression of the fermionic spectral function with strong spatial modulation. In particular, when the amplitude of the ionic lattice becomes sufficiently large, the Fermi surface along the direction of broken translational symmetry gradually disappears, leaving behind \emph{disconnected segments}, as visible from figure~\ref{fig:densityq2ionic} 
(in contrast, the spectral function along the direction that respects translational invariance seems to be enhanced). 
The behavior of the Fermi surface is reminiscent of the spectral signatures observed in modulated 
superconducting phases and the 
discussion of open Fermi surface segments and Fermi arcs appearing in PDW phases, see \emph{e.g.}~\cite{Baruch:2008,Berg:2009,Lee:2014,Garrido:2015}.
In particular, we anticipate that in the fully crystallized case (in which translational invariance is broken along all boundary directions) the Fermi surface will consist of detached pieces similar to Fermi arcs
(see also the discussion of Fermi surface reconstruction in \emph{e.g.}~\cite{Sachdev:2012if}). 
We suspect that the feature we have identified -- the gradual disappearance of the Fermi surface with increasing lattice strength -- 
is a general property in holography and may not be very sensitive to the specific type of spatial modulation in the system. 
We have seen preliminary evidence of 
this general behavior in~\cite{CurrentWork} and plan to report on it in follow-up work\,\footnote{Notice that 
disorder can also lead to the gradual disappearance of Fermi arcs (see \emph{e.g.} the effects described in~\cite{Roy}).
Indeed, disorder can also pin density wave order, but such pinning is random in nature. Thus, the role of disorder and lattice potentials on density-wave order is different.}.\\

To confirm our intuition and determine whether this phenomenon is a generic signature of strong translational symmetry breaking, we need to examine the fermionic response in  
additional classes of models\,\footnote{The PDW model of~\cite{Cai:2017qdz}, which involves parity breaking terms, is similar in spirit to ours.}, with and without spontaneously generated stripe order.
%%%%
For example, in subsection~\ref{appC} we have examined a simpler model which includes an explicit ionic lattice but no spontaneously generated stripes.
In this setup we also observe the segmentation of the Fermi surface  with a strong enough ionic lattice. 
It is natural to ask if the same effect would happen by solely increasing the spontaneous modulation (an IR effect), or whether 
it is only due to the UV lattice.
While we naively expect that a sufficiently strong PDW would also lead to the suppression of the spectral weight, in order to show it 
the background geometry needs to be constructed at a much lower temperature. Moreover, recall that in this analysis we have considered the simple case in which the period of the spontaneously generated stripes is commensurate with the lattice spacing -- corresponding to a single length scale in the system. 
Thus, in our construction the ionic lattice potential acts to amplify the effects of the PDW, making it difficult to 
 disentangle the specific role played by the different symmetry breaking mechanisms (spontaneous and explicit). 
To clarify this point it would be interesting to generalize the present study to the incommensurate case, in which the two periods describe independent physical scales. 
We expect the incommensurability to lead to novel effects.
 
 While our analysis clearly shows that the Fermi surface dissolves at large lattice strength, 
 we are still lacking a deeper understanding of the origin of this phenomenon, and of
its role in the context of high temperature superconductors.
In particular, it would be valuable to use holographic studies such as ours to distinguish between the possible scenarios put forth to explain the appearance of Fermi surface segments -- for instance, real Fermi arcs, point nodes and a small gap at the nodal point would all appear as arcs due to thermal broadening (see \emph{e.g.}~\cite{Vishik}). 
We note that in our analysis thus far we don't see evidence for the suggestion that these arcs could be segments of Fermi pockets. 
It would be interesting to be able to rule out this possibility, and more importantly to isolate  
specific predictions of our model which could potentially be reproduced  by experiment.

Another relevant question is that of the behavior of the low energy excitations near the Fermi surface. In our holographic construction of PDW phases, spatial translations are broken spontaneously, resulting in a strongly relevant spatially modulated deformation in the IR. It is well known that the radial direction in the bulk plays the role of the energy scale in the dual system, and that excitations with different wavelengths are mapped to different regions of the bulk. Thus, the low energy behavior around the Fermi surface 
could be traced back to the near horizon geometry of the bulk configuration. 
We stress that the low energy physics in the spatially modulated phase due to spontaneous translational symmetry breaking will be starkly different from that in an homogeneous background or in the case with an irrelevant lattice, for which the low temperature IR geometry reduces to the homogeneous one. 

While in this paper we have restricted our attention to a free bulk fermion, one can couple it to various intertwined orders, considering \emph{e.g.} a Majorana coupling~\cite{Faulkner:2009am} and dipole interaction~\cite{Edalati:2010ww,Li:2011nz}. 
By coupling the fermion to the scalar in our model 
it may be possible to reproduce some of the spectral weight features seen in the mean-field theory analysis of~\cite{Baruch:2008}, in which arcs in momentum space shrunk with increasing superconducting order.
Finally, in the present analysis we have limited ourselves to the low (but finite) temperature case, partially due to the limitations of our computing resources. 
Although the computation becomes more challenging, it would be interesting to construct the background geometry as $T\rightarrow 0$ and study the associated fermionic response in the extremal case, thus providing a window into the structure of the 
ground state. We leave these questions to future work.

\begin{acknowledgments}
We are grateful to Eduardo Fradkin for many insightful conversations and to Bitan Roy for comments on the draft. 
We also thank Alexander Krikun, Yan Liu, Chao Niu, Koenraad Schalm and Jan Zaanen for helpful discussions. 
The work of S.C. is supported in part by the National Science Foundation grant PHY-1620169.
J.R. is partially supported by the American-Israeli Bi-National Science Foundation, the Israel Science Foundation Center of Excellence and the I-Core Program of the Planning and Budgeting Committee and The Israel Science Foundation ``The Quantum Universe."
\end{acknowledgments}

\newpage
\appendix
\renewcommand{\theequation}{\thesection.\arabic{equation}}
\addcontentsline{toc}{section}{Appendix}
%\section*{Appendices}

\section{Energy Distribution of the Spectral Density}
\label{appA}

We examine the behavior of the energy distribution function (EDF), i.e. the spectral density as a function of $\omega$, for different choices of momentum. 
We are specifically interested in how the spectral density evolves as the momentum is varied, in order to probe the formation and evolution of the Fermi surface itself.
These results provide an additional check of the analysis presented in the main text and highlight some interesting features.

\subsection{Case (i): PDW without Ionic Lattice}

We focus on the features of the Fermi surface we noted in figure~\ref{fig:densityq2} and consider two interesting cases:  

\begin{itemize}
  \item (1) we fix $k_x=0.8$  (which places us near the boundary of the Brillouin zone) and increase $k_y$ so that it crosses the inner and outer Fermi surfaces;
  \item (2) we fix $k_x=0$ (which places us at the center of the Brillouin zone) 
  and vary $k_y$ so that it crosses the bump and the outer Fermi surface.
\end{itemize}

\begin{figure}[ht!]
\begin{center}
\includegraphics[width=.45\textwidth]{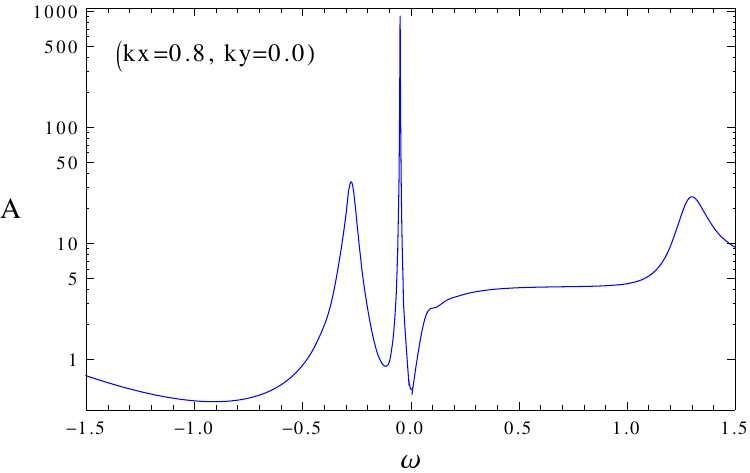}\quad
\includegraphics[width=.45\textwidth]{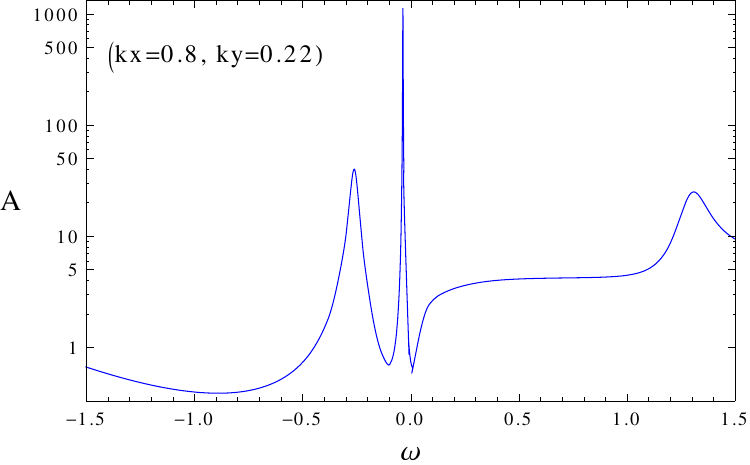}
\includegraphics[width=.45\textwidth]{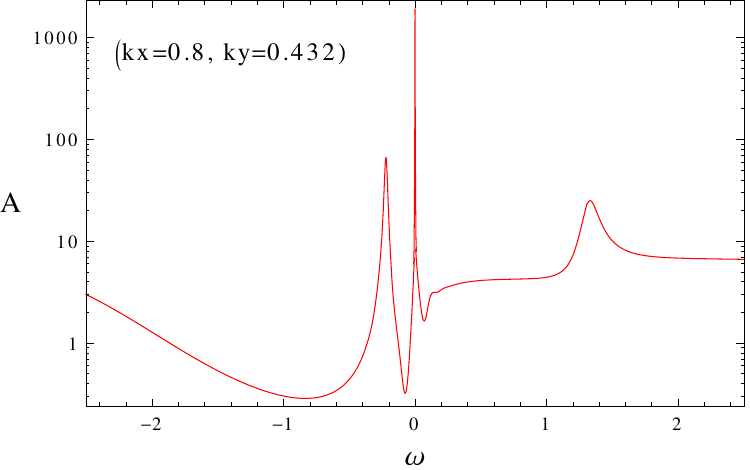}\quad
\includegraphics[width=.45\textwidth]{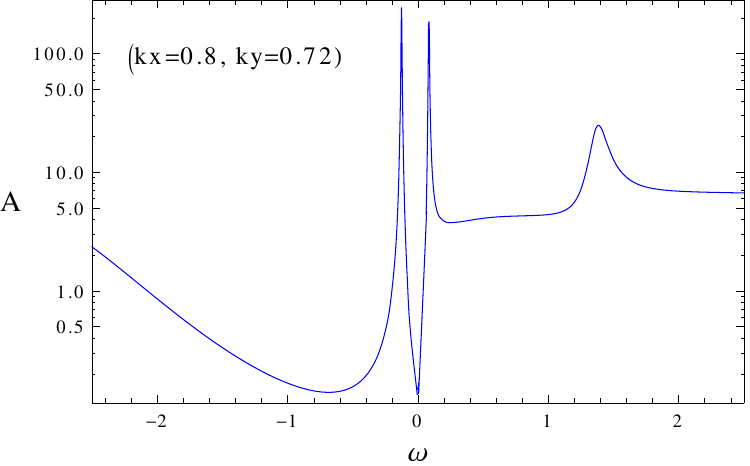}
\includegraphics[width=.45\textwidth]{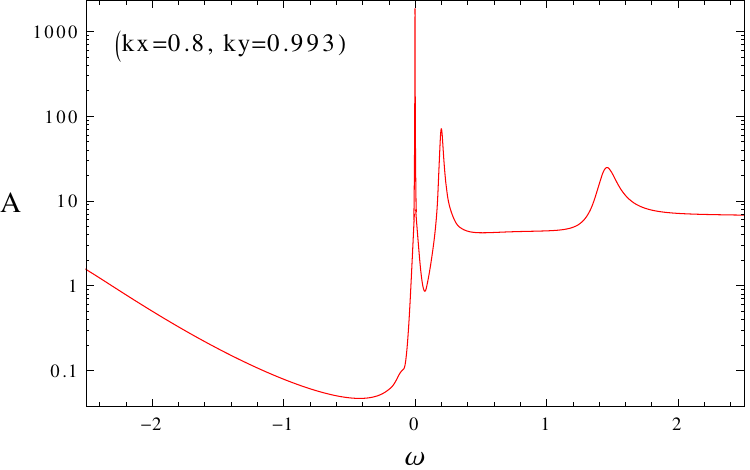}\quad
\includegraphics[width=.45\textwidth]{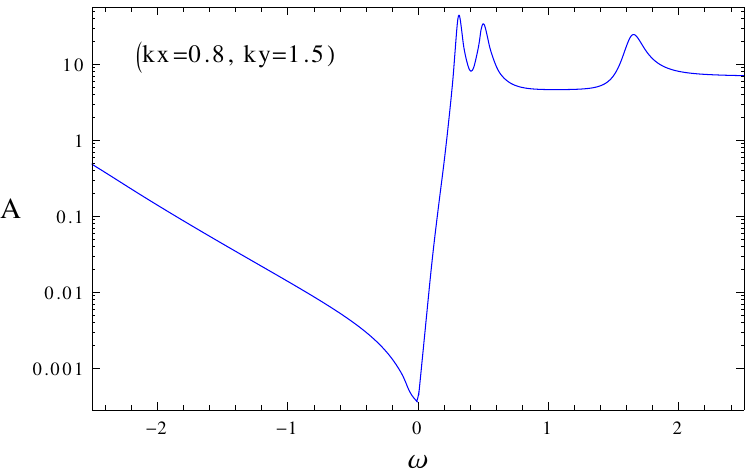}
\caption{Behavior of the EDF at different points in the momentum plane. 
We fix $k_x=0.8$ and increase $k_y$ from points inside the inner Fermi surface to points outside the outer Fermi surface. The two plots in red denote cases in which a Fermi surface is present.
Note that the vertical axis is logarithmic. We have used the same setup as in figure~\ref{fig:densityq2}.}
\label{fig:Avswkx08}
\end{center}
\end{figure}

Typical behaviors of the EDF for the first case (near the edge of the Brillouin zone at $k_x=0.8$) are presented in figure~\ref{fig:Avswkx08}.
When $k_y=0$ we see a small broad peak and a large sharp peak, both on the left side of the vertical axis $\omega=0$. 
We stress that the location of the sharp peak is away from $\omega=0$, and thus it should not be regarded as evidence for a Fermi surface. As we increase $k_y$, the sharp peak moves to the right with its amplitude also increasing, getting closer to the $\omega$-axis.
We find a very sharp peak at $\omega=0$ when $k_y\approx0.432$ (third panel in figure~\ref{fig:Avswkx08}), at which point the inner Fermi surface develops. 
As we continue increasing the value of $k_y$, the sharp peak moves to the positive $\omega$ axis and becomes smooth and small.  
In contrast, as $k_y$ is increased the broad peak moves closer to the $\omega$-axis and becomes sharper and larger. 
It moves to the $\omega=0$ axis when $k_y\approx0.993$ (fifth panel in figure~\ref{fig:Avswkx08}), which identifies the location 
of the outer Fermi surface. If we increase $k_y$ further, this peak moves to the right of the $\omega$-axis and becomes smooth. This confirms what we see in figure \ref{fig:densityq2}.
We do not see any evidence of Fermi surface for $k_y>0.993$. Instead, we find that the values of the spectral density $A$ near $\omega=0$ for large $k_y$ are quite small.

\begin{figure}[ht!]
\begin{center}
\includegraphics[width=.45\textwidth]{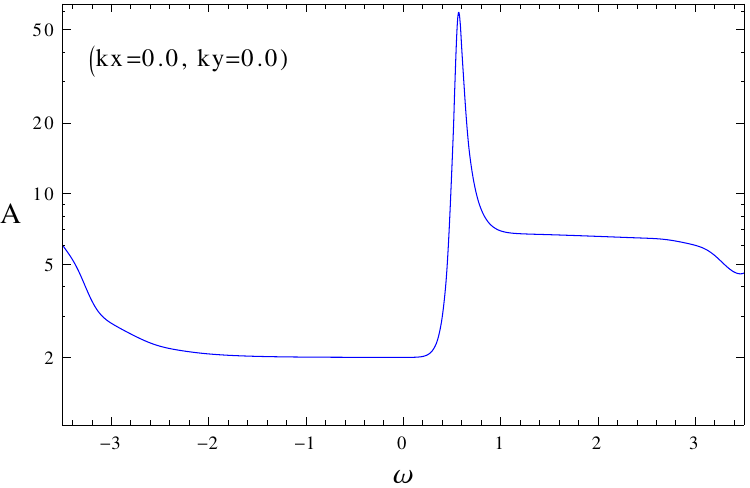}\quad
\includegraphics[width=.45\textwidth]{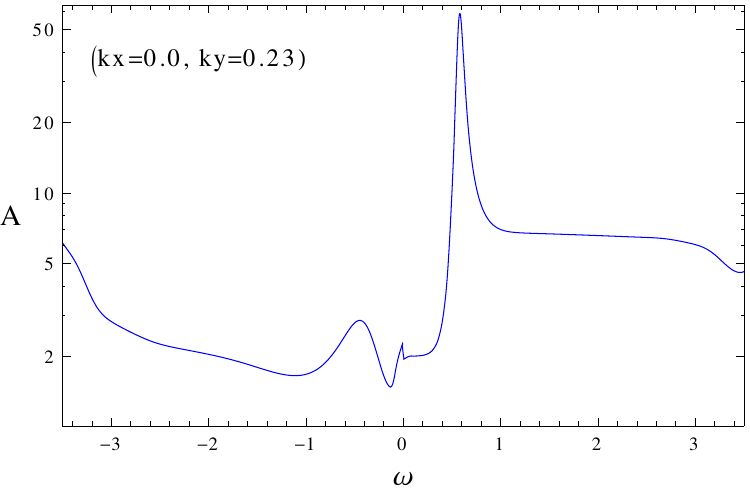}
\includegraphics[width=.45\textwidth]{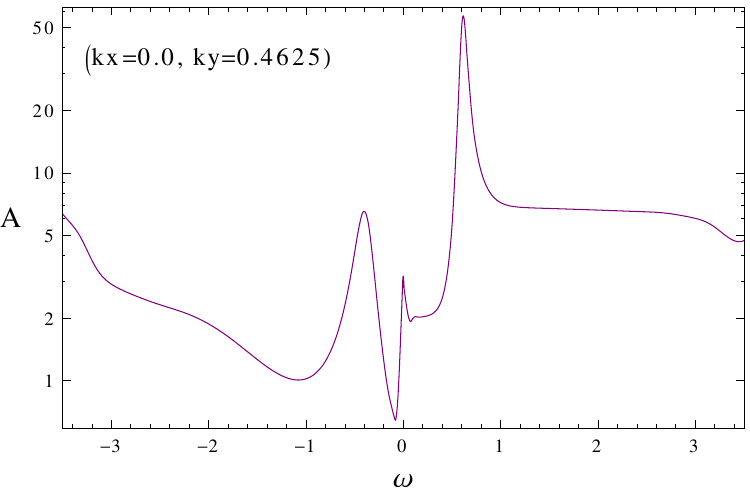}\quad
\includegraphics[width=.45\textwidth]{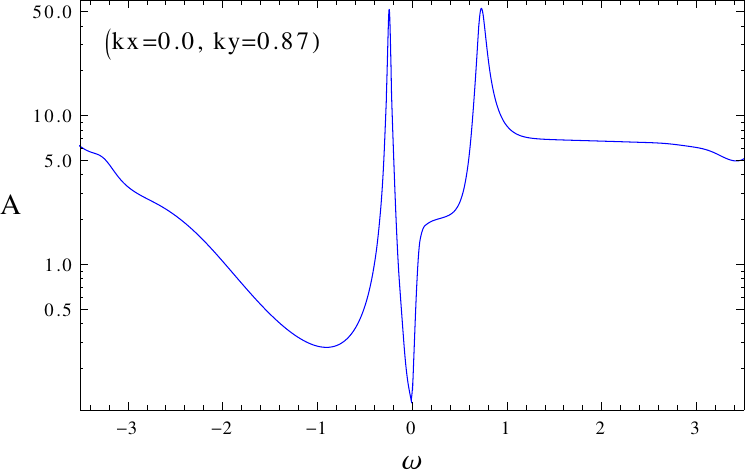}
\includegraphics[width=.45\textwidth]{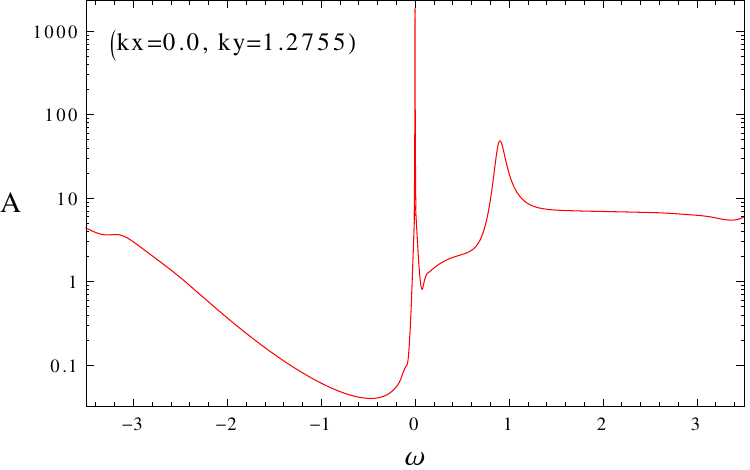}\quad
\includegraphics[width=.45\textwidth]{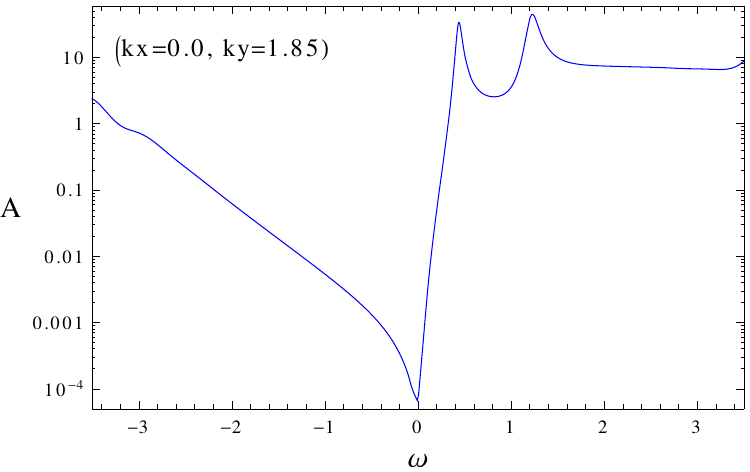}
\caption{Spectral density $A$ as a function of $\omega$ at different points in the momentum plane.
We fix $k_x=0.0$ and increase $k_y$ from $0.0$ to $1.85$.
 The plot marked in red corresponds to the presence of a Fermi surface, and the one marked in purple shows the development of a small peak at $\omega=0$.  Note that the vertical axis is logarithmic. We choose the same parameters as in figure~\ref{fig:densityq2}.}
\label{fig:Avswkx00}
\end{center}
\end{figure}

We now turn to the second case for which $k_x=0$, i.e. in the middle of the Brillouin zone. Representative plots are shown in figure~\ref{fig:Avswkx00}. The behavior at $k_y=0$ is quite simple -- there is only one smooth peak at $\omega\approx0.58$ (see the first panel in the figure). As the value of $k_y$ increases, two more small peaks appear near the $\omega$-axis, with their amplitudes growing. A peak develops at $\omega=0$ when $k_y\approx0.4625$, with a small amplitude and a broad width, such that it should not be regarded as a Fermi surface. 
It corresponds instead to one of the points on the small circle of broad peaks in figure~\ref{fig:densityq2}. As we continue increasing $k_y$, the successive peak shifts toward the $\omega$-axis and becomes sharper and sharper. The Fermi surface finally forms at $k_y\approx 1.2755$ (fifth panel in the figure). For larger values of $k_y$  there are no additional sharp peaks developing at $\omega=0$.

Another feature we observe which appears to be in agreement with ARPES experiments on the cuprates is the ``peak-dip-hump" structure in the EDF at fixed momentum (see \emph{e.g.} ~\cite{PDH1,PDH2,PDH3,PDH4,PDH5}). 
Indeed, in the first two panels of figure~\ref{fig:Avswkx08} one sees a sharp low energy peak accompanied by a broad maximum at larger values of $\omega$, reminiscent of what is observed
in the spectrum of several high-$T_c$ superconductors. For previous holographic models discussing this feature see~\cite{Chen:2009pt,Faulkner:2009am}.
We would like to examine this structure in further detail, in part because of its potential relation to laboratory systems, and understand 
its origin and whether it is generic.

\subsection{Case (ii): PDW with Ionic Lattice}

\begin{figure}[ht!]
\begin{center}
\includegraphics[width=.45\textwidth]{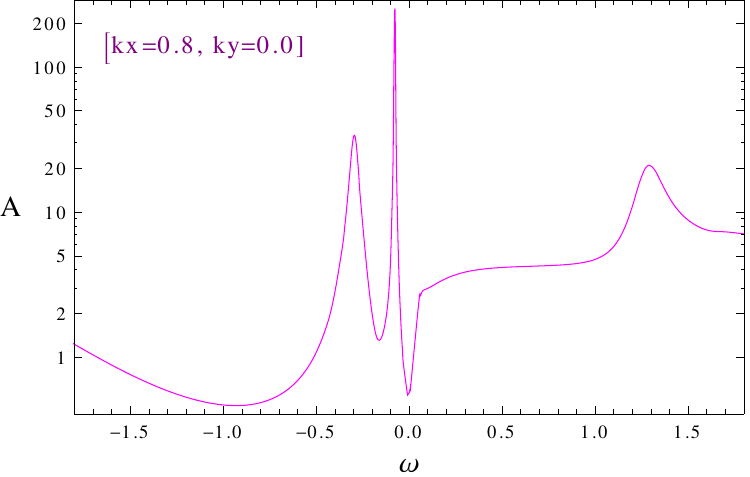}\quad
\includegraphics[width=.45\textwidth]{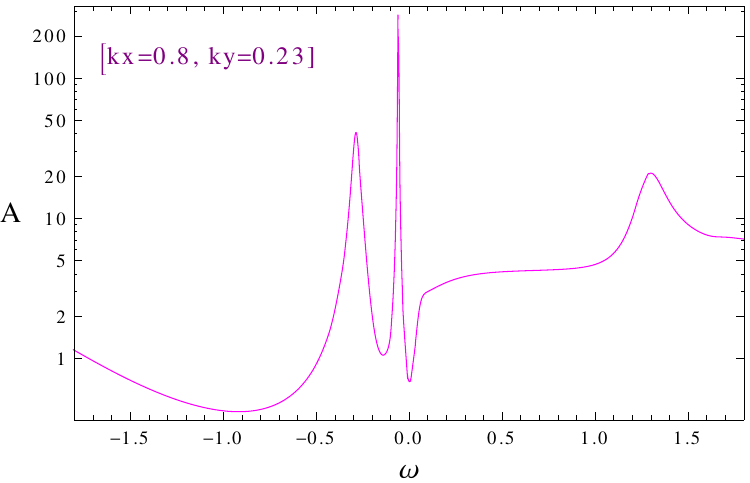}
\includegraphics[width=.45\textwidth]{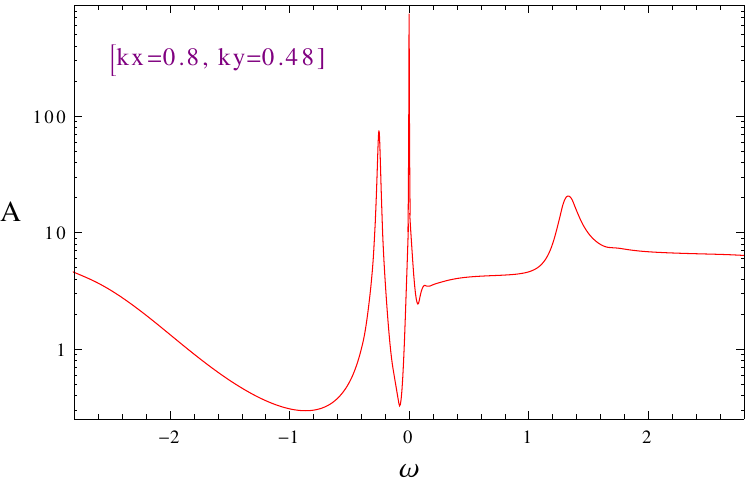}\quad
\includegraphics[width=.45\textwidth]{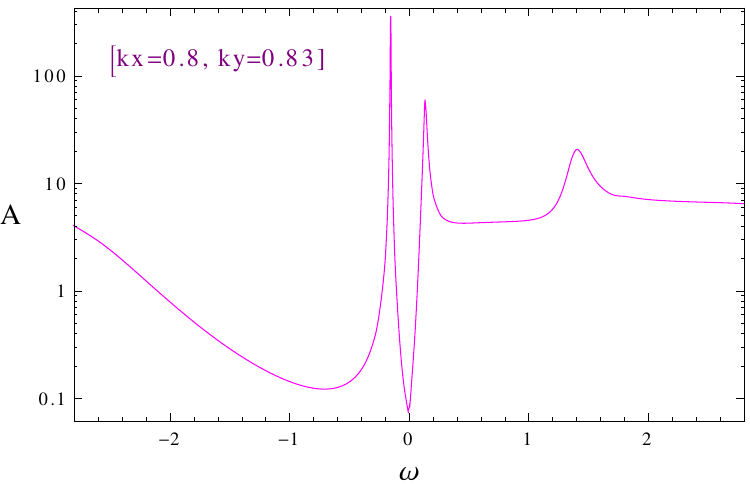}
\includegraphics[width=.45\textwidth]{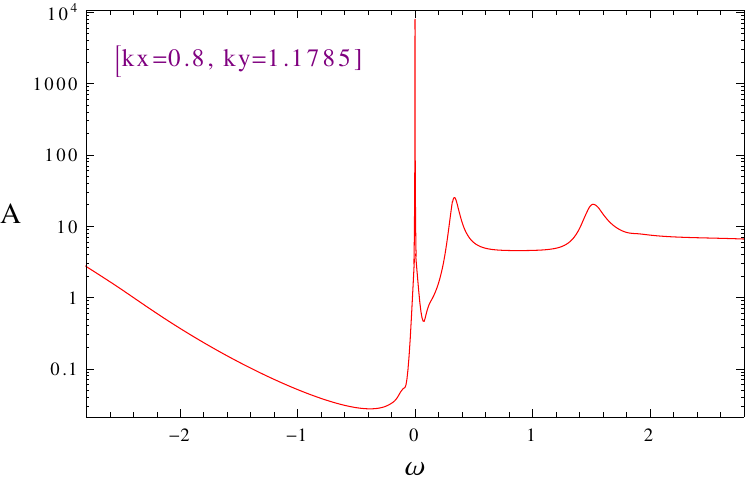}\quad
\includegraphics[width=.45\textwidth]{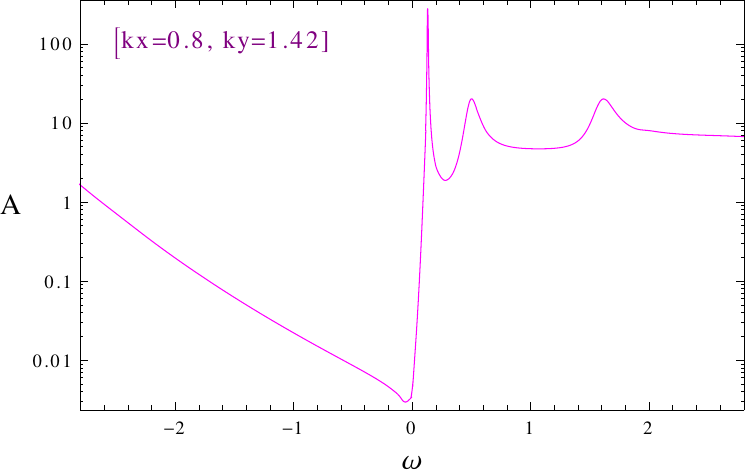}
\caption{The behavior of the EDF at different momenta for the PDW phase with the ionic lattice. The two plots exhibiting Fermi surfaces are marked in red. We fix $k_x=0.8$ and increase $k_y$ from points inside the inner Fermi surface to points outside the outer Fermi surface. Note that the vertical axis is logarithmic. We have used the same setup as in figure~\ref{fig:densityq2ionic}.}
\label{fig:Avswkx08ionic}
\end{center}
\end{figure}

The behavior of the spectral density as a function of $\omega$ in the presence of an explicit lattice
is quite similar to that of the pure PDW case. As a comparison, in figure~\ref{fig:Avswkx08ionic} we show our results with $k_x=0.8$ as well as a similar choice of $k_y$ as that of figure~\ref{fig:Avswkx08}.
The presence of a Fermi surface is now visible in the third and fifth panels of figure~\ref{fig:Avswkx08ionic}. 
The first two panels reveal the same kind of peak-dip-hump structure we noticed in the pure PDW case.

\section{Details of Numerical Analysis}
\label{appB}

Due to the absence of analytic solutions for holographic striped superconductors, we have employed numerical techniques to solve the PDEs and calculate the fermionic spectral density. 
The solution for the background geometries is described in section~\ref{backreaction} and in our earlier work~\cite{Cremonini:2017usb}. As shown in Appendix~D of \cite{Cremonini:2017usb}, the accuracy of our numerical calculations was checked in two ways. The first one is the convergence of $\xi^2$ as the grid size is increased, and the second one is the first law of thermodynamics.

\begin{figure}[ht!]
\begin{center}
\includegraphics[width=.45\textwidth]{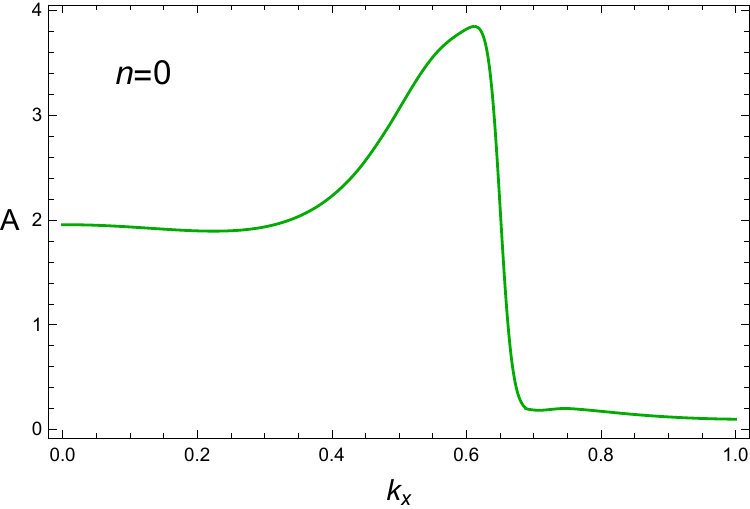}\quad\quad
\includegraphics[width=.45\textwidth]{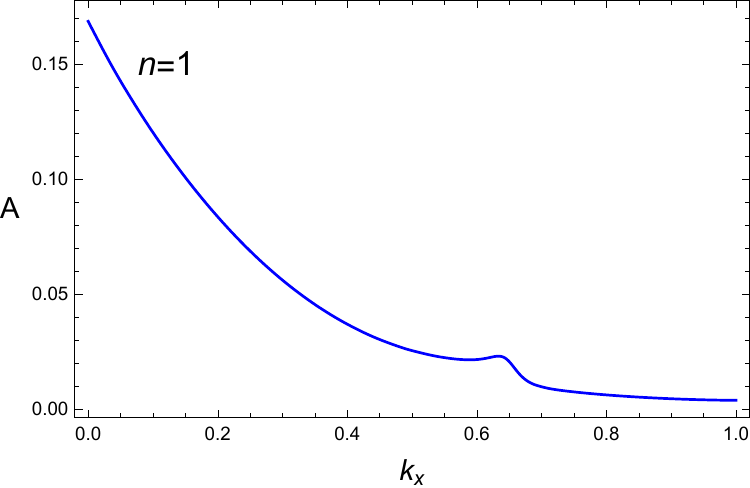}\\
\includegraphics[width=.45\textwidth]{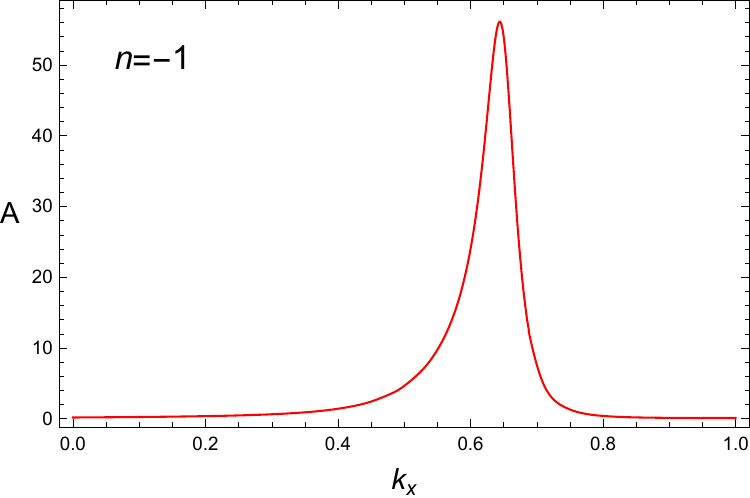}\quad\quad
\includegraphics[width=.45\textwidth]{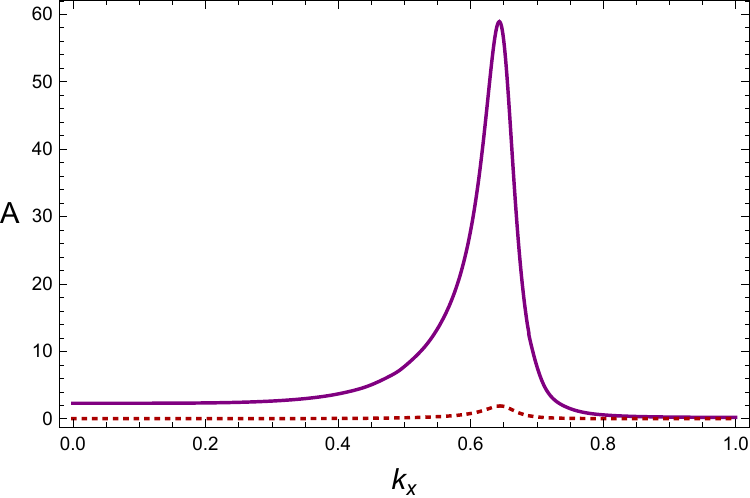}
\caption{The contribution to the spectral function for each diagonal component of retarded Green's functions. The first three plots correspond to $n=0$, $n=1$ and $n=-1$, espectively. They all have a smooth peak near $k_x=0.65$.  In the fourth plot the purple curve is the sum of these three modes, while the dotted curve corresponds to the sum of the four higher modes with $n=\pm 2$ and $n=\pm 3$. We choose $\omega=10^{-6}, k_y=0$ and $q=2$. The background geometry is the same as in figure~\ref{fig:geometryIonic} with the ionic lattice amplitude $a_0=0.5$ and the wavenumber $p=2$.}
\label{fig:modes}
\end{center}
\end{figure}

The Green's function $G_{\alpha,n;\,\alpha',n'}(\omega,k_x,k_y)$ is obtained after solving the Dirac equation. 
To solve the latter we have used the pseudo-spectral collocation approximation to convert the PDEs~\eqref{Dirac1} and~\eqref{Dirac2} into linear algebraic equations, by adopting Fourier discretization in the $x$ direction and Chebyshev polynomials in the $z$ direction. Note that the introduction of the coordinate systems~\eqref{ansatzbh} ensures that the solutions are smooth near the horizon. 
Thus, we can use a relatively smaller number of grid points to solve the system, thanks to the pseudo-spectral method. When solving the equations,
we can either use the same grid size as the background solution, or interpolate from the background solution. Both the background geometry and the Dirac equation were independently solved by two of the authors and compared for agreement.

Note that we have defined the spectral function $A$ in~\eqref{spectral} by summing the imaginary part of the diagonal components of the retarded Green's function. 
Since we have explicitly checked that the contribution from higher modes with large $n$ is very small, we can neglect them when we compute $A$. 
A typical result of our analysis is presented in figure~\ref{fig:modes}, where we show the contribution from different diagonal components. 
First of all, note that the behavior of each diagonal component is quite different, indicating that it is crucial to identify the dominant contributions and sum over all of them. 
In our case the dominant contribution comes from the first three modes with $(n=0, n=\pm 1)$. The contribution from higher modes is too small to change the behavior, 
as shown by the dotted line in the fourth plot of figure~\ref{fig:modes}. 
We also checked that the contribution from higher modes becomes smaller and smaller as the amplitude of the ionic lattice is decreased.
Thus, from our analysis we see that it is a good approximation to consider the lowest three modes with $(n=0, n=\pm 1)$ when computing the spectral function $A$. 
For the cases we considered in the main text, this is sufficient to see the suppression of the spectral weight as one increases the amplitude of the ionic lattice.

\begin{figure}[ht!]
\begin{center}
\includegraphics[width=.50\textwidth]{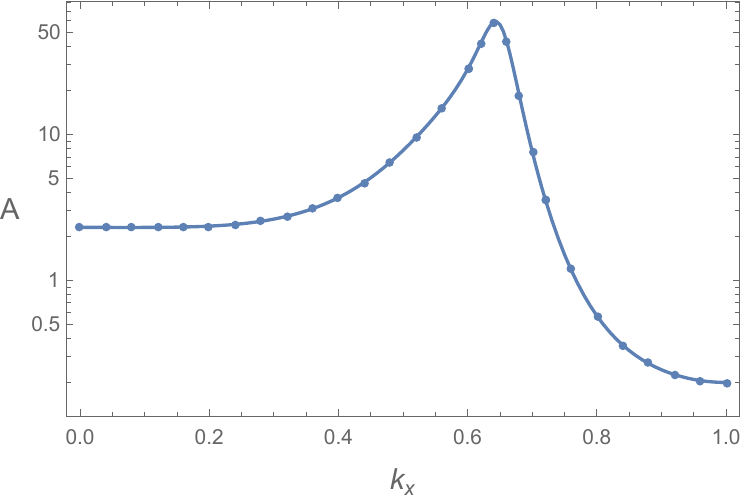}
\caption{The MDF along the $k_x$-axis for $a_0=0.5$, $\omega=10^{-6}$, $q=2$ and $T=0.01426$. The curve is computed by choosing $N_z=N_x=21$, and the dots are obtained by $N_z=N_x=51$ (the grid size for the background geometry).}
\label{fig:size21}
\end{center}
\end{figure}

For the PDW without ionic lattice, we used the same grid size for both the background solution and the Dirac equation. The grid size is $N_z=N_x=31$ when we plot the temperature dependence as in figure~\ref{fig:gapvsT}, and is $N_z=N_x=21$ when we plot the 3D spectral density as in figure~\ref{fig:specvsky} (as well as the density plot in figure~\ref{fig:densityq2}). For the PDW with ionic lattice, we have to increase the grid size as we increase the amplitude $a_0$. We have used $N_z=N_x=51$ to obtain the background geometries. We interpolate to a relatively smaller size $N_z=N_x=21$ to obtain figure~\ref{fig:densityq2ionic}. For the other figures we have used a slightly larger size of grid points, $(N_z, N_x)\sim 35$. We have checked that our choice of grid size does not change the spectral density in a visible way, as shown in figure~\ref{fig:size21}.

\end{document}